\documentclass[aps,pra,twocolumn,groupedaddress,nofootinbib,longbibliography]{revtex4-2}
\usepackage[utf8]{inputenc}
\usepackage[english]{babel}
\usepackage[T1]{fontenc}
\usepackage{amsfonts}

\usepackage{hyperref}

\usepackage{tikz}

\usepackage{mathtools}
\usepackage{braket}
\usepackage[charter,cal=cmcal,sfscaled=false]{mathdesign}
\definecolor{iffsred}{cmyk}{0.12,0.94,0.87,0.34}
\definecolor{uestcblue}{cmyk}{0.99,0.78,0.16,0.03}

\hypersetup{
  pdftitle={Contextuality as a Diagnostic of Translation-Symmetry Breaking in Translation-Invariant 1D Hamiltonians},
  pdfauthor={Zeng, Yang, Zhang and Wang.},
  pdfstartview=Fit,
  pdfpagelayout=SinglePage,
  colorlinks,
  linkcolor=uestcblue,
  citecolor=uestcblue,
  urlcolor=iffsred}

\newcommand{\TI}{\mathrm{TI}}
\newcommand{\PBC}{\mathrm{PBC}}
\newcommand{\OBC}{\mathrm{OBC}}

\begin{document}

\title{Contextuality as a Diagnostic of Translation-Symmetry Breaking in Translation-Invariant 1D Hamiltonians}

\author{Xiao Zeng}
\affiliation{Institute of Fundamental and Frontier Sciences, University of Electronic Science and Technology of China, 611731, Chengdu, China}
\affiliation{Ministry of Education Key Laboratory of Quantum Physics and Photonic Quantum Information (University of Electronic Science and Technology of China), 611731, Chengdu, China}
\author{Kaiyan Yang}
\author{Lingxia Zhang}
\affiliation{Institute of Fundamental and Frontier Sciences, University of Electronic Science and Technology of China, 611731, Chengdu, China}
\affiliation{Ministry of Education Key Laboratory of Quantum Physics and Photonic Quantum Information (University of Electronic Science and Technology of China), 611731, Chengdu, China}
\author{Zizhu Wang}\email{zizhu@uestc.edu.cn}
\affiliation{Institute of Fundamental and Frontier Sciences, University of Electronic Science and Technology of China, 611731, Chengdu, China}
\affiliation{Ministry of Education Key Laboratory of Quantum Physics and Photonic Quantum Information (University of Electronic Science and Technology of China), 611731, Chengdu, China}

\begin{abstract}
Bell- and contextuality-type inequalities have become practical probes of many-body quantum correlations, often involving only few-body correlators and quantities with a direct Hamiltonian interpretation such as an energy density. Here we investigate the mechanism by which translation-invariant Hamiltonians generated from representative families of contextuality witnesses organize their ground-state structure in infinite one-dimensional systems. For the witness families considered, maximal quantum violation is realized by ground-state sectors with commensurate enlarged unit cells: the Hamiltonians are invariant under one-site translations, while the optimal ground states are $p$-periodic with $p>1$. At the corresponding classical-bound points, the ground-state sectors are highly degenerate and support many commensurate periods. Along the interpolation paths analyzed in this work, entering the contextual regime is accompanied by the lifting of this classical period degeneracy in favor of a quantum-selected period. We also identify finite periodic-boundary-condition benchmarks at the selected periods: for each model studied, the finite-ring witness reproduces the same classical bound and quantum value as the corresponding infinite-chain witness, and in several cases the resulting finite inequalities are tight. These reductions turn the infinite-chain contextuality certification into compact energy-estimation benchmarks requiring only local correlator measurements. We establish the mechanism analytically in representative two- and three-body witness models and corroborate it more broadly using translation-invariant semidefinite-program relaxations together with variational matrix-product-state calculations.
\end{abstract}
\maketitle

\section{Introduction}

Bell's theorem establishes that quantum correlations can be incompatible with any local hidden-variable model, and it provides an operational route---Bell inequalities---to certify this incompatibility from measured statistics \cite{bellEinsteinPodolskyRosen1964,nicolasbrunnerBellNonlocality2014}.
In many-body settings, however, strict spacelike separation between subsystems is typically unavailable; violations of Bell-type inequalities are then interpreted as signatures of \emph{contextuality} rather than genuine spatial nonlocality \cite{kochen1967problem,budroniKochenSpeckerContextuality2022,cabelloBellNonlocalityKochenSpecker2021}.
Beyond foundational significance, contextuality has also been identified as a resource underpinning quantum advantages in information processing \cite{howardContextualitySuppliesMagic2014,bermejo-vegaContextualityResourceModels2017,bravyiQuantumAdvantageShallow2018,bravyiQuantumAdvantageNoisy2020,zurelHiddenVariableModel2020,guptaQuantumContextualityProvides2023a}.

A parallel line of work has developed \emph{many-body} Bell and contextuality witnesses built from few-body correlators and tailored to the symmetries and measurement constraints of correlated-matter platforms.
Notably, translationally invariant (TI) multipartite inequalities can be constructed using only one- and two-body correlators \cite{turaDetectingNonlocalityManybody2014a,jturaTranslationallyInvariantMultipartite2014,hu2025characterizingtranslationinvariantbellinequalities,hu2024tropicalcontractiontensornetworks}, and the expectation value of such a witness can often be interpreted as the energy density of a corresponding local Hamiltonian \cite{j.turaEnergyDetectorNonlocality2017}.
This ``energy-as-witness'' viewpoint is particularly appealing experimentally: local energies and short-range correlators are among the most accessible observables in quantum simulators and processors.

Translation invariance becomes even more powerful in the thermodynamic limit, where one can ask whether the reduced state of $m$ neighboring sites is compatible with an infinite TI extension and whether its correlators can violate a TI classical model \cite{wangEntanglementNonlocalityInfinite2017,fineHiddenVariablesJoint1982}.
In the thermodynamic limit, this energy-as-witness viewpoint turns a question about contextuality into a ground-state problem for a translation-invariant local Hamiltonian. The witness value is an energy density, and the optimal quantum violation is determined by the ground state selected by the corresponding Hamiltonian. This raises a condensed-matter question: what type of many-body order, if any, is associated with the Hamiltonians that reach the maximal contextuality violation?

In this work, we show that, within the witness families studied, the distinguishing many-body principle behind maximal violation is \emph{macroscopic commensurate order} arising from spontaneous breaking of one-site translation symmetry. Although the Hamiltonians are invariant under single-site translations, their ground spaces can enlarge the unit cell and become strictly 
$p$-periodic ($p>1$), with a degenerate manifold of ground states related by lattice translations. Such translation-symmetry breaking is a canonical organizing mechanism in one-dimensional spin chains. For instance, at the Majumdar--Ghosh point the spin-$1/2$ $J_1$--$J_2$ Heisenberg chain has exactly dimerized ground states with a two-site unit cell \cite{Majumdar:1969zmb,Majumdar1969OnNI}, and competing-interaction Heisenberg chains exhibit well-characterized fluid--dimer transitions \cite{Haldane1982,OKAMOTO1992433}. More broadly, competing interactions can stabilize families of commensurate modulated phases with distinct periods, as in the axial next-nearest-neighbor Ising (ANNNI) model and its “devil’s staircase” phenomenology \cite{FisherSelke1980,Bak1982,Selke1988}. Finally, commensurability constraints of Lieb--Schultz--Mattis type imply that gapped phases such as magnetization plateaus may require an enlarged unit cell, enforcing translation-symmetry breaking or ground-state degeneracy \cite{LiebSchultzMattis1961,OshikawaYamanakaAffleck1997,Oshikawa2000,Hastings2004}.

\begin{figure}[htbp]
\centering
\includegraphics[width=8.5cm]{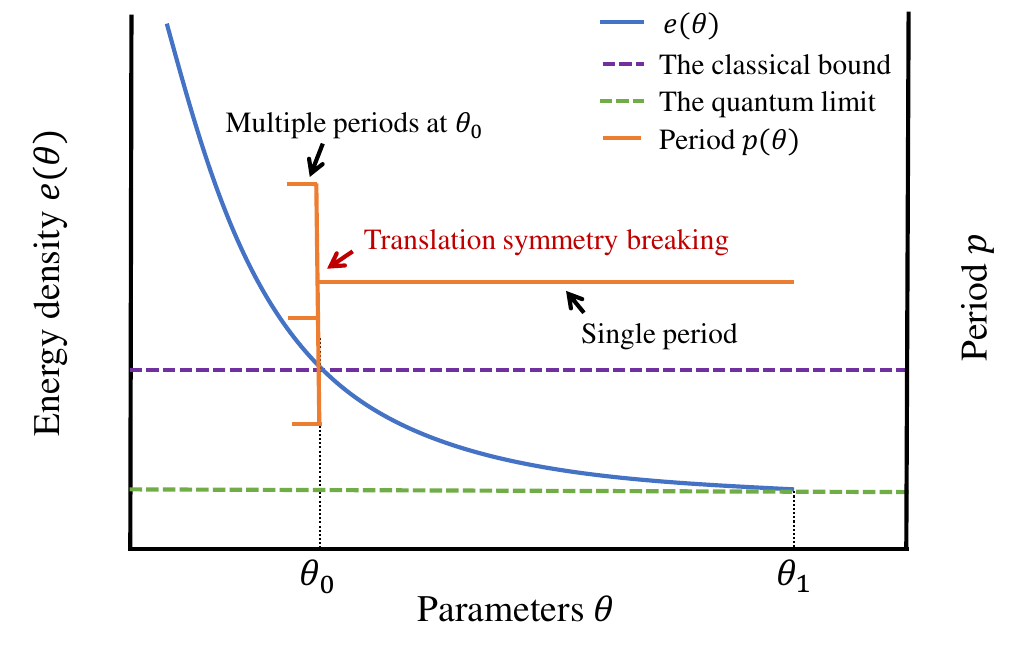}
\caption{Schematic classical-to-quantum symmetry selection along a parameterized Hamiltonian family $H(\theta)$. We assume $H(\theta_0)$ saturates the classical bound while $H(\theta_1)$ reaches the maximal quantum violation. As $\theta$ is tuned from $\theta_0$ to $\theta_1$, the ground-state energy density $e(\theta)$ decreases and contextuality appears. At $\theta_0$, the classical-saturating Hamiltonian has a ground space supporting multiple distinct commensurate periods. For $\theta>\theta_0$, this period degeneracy is lifted and the ground state becomes strictly $p_{\mathcal{Q}}$-periodic with $p_{\mathcal{Q}}>1$, signaling spontaneous breaking of one-site translation symmetry.}
\label{conceptual}
\end{figure}

Our main physical message can be stated succinctly: for our translation-invariant contextuality witness families, the maximal contextuality is consistently accompanied by commensurate macroscopic order.
Specifically, we identify a sharp organizing principle for the
extremal classical and quantum strategies in the thermodynamic limit. 
(i) Hamiltonians that \emph{saturate the classical bound} exhibit an extensively degenerate ground space that supports many distinct commensurate periods.
(ii) By contrast, Hamiltonians that \emph{attain the quantum optimum} select a unique strictly periodic ground-state structure of period $p_{\mathcal{Q}}>1$ up to translations, implying spontaneous breaking of one-site translation symmetry.
Along natural continuous interpolations between these endpoints, the classical bound acts simultaneously as the contextuality threshold and the symmetry-selection point: in the interpolations studied, entering the contextual regime coincides with lifting the classical period degeneracy in favor of a quantum-selected period.
A schematic illustration is shown in Figure~\ref{conceptual}.

Moreover, for each witness studied, we identify a finite PBC benchmark at
the quantum-selected period $p_{\mathcal{Q}}$. The induced $p_{\mathcal{Q}}$-partite periodic-boundary-condition (PBC) inequalities on a ring have the same classical and quantum bounds as their infinite translation-invariant counterparts. In several cases, the resulting PBC inequalities are tight, in the sense that they define facets of the TI classical polytope on the $p_{\mathcal{Q}}$-site ring. This elevates our TI contextuality witnesses from abstract certificates of nonclassicality to symmetry-sensitive \emph{order diagnostics} with experiment-ready signatures—compact ring benchmarks implementable on quantum processors. We establish the mechanism analytically in representative 232- and 322-type models and corroborate it more broadly using TI-adapted semidefinite-program certificates, which agree with variational uniform matrix-product-state (MPS) calculations to solver precision. Separately, we show that a finite 232-type witness admits valid open-boundary-condition (OBC) realizations whose corresponding OBC Hamiltonians yield robust violations in numerics up to 100 parties.

The remainder of the paper is organized as follows.
Section~\ref{sec:background} reviews background on the translation-invariant witness families studied, their Hamiltonian representation, and diagnostics of translation-symmetry breaking based on MPS transfer matrices.
Section~\ref{sec:main-results} presents the main results and analyzes representative models—including exactly solvable cases—to illustrate the emergence of translation-symmetry breaking at maximal contextuality.
Section~\ref{sec:classical} explains the degeneracy mechanism at the classical bound and its relation to TI classical models via domino-loop constructions.
Section~\ref{sec:npa} develops a TI-adapted semidefinite-program framework,
constructs the finite PBC reductions, and derives analytical solutions for representative 322-type models.
Section~\ref{sec:benchmark} formulates practical benchmarking protocols for quantum processors and simulators, together with their measurement signatures.
We conclude in Section~\ref{sec:discussion} with implications and an outlook.

\section{Background}\label{sec:background}
This section briefly reviews the background relevant to this work, including the TI contextuality
witness families studied, the Hamiltonian formulation, and diagnostics of translation-symmetry breaking based
on MPS transfer matrices.
\subsection{Setting: translation-invariant contextuality witnesses as Hamiltonians}
In this work, we consider two types of TI contextuality witnesses involving only one- and two-body correlators: the 232- and 322-type witnesses, originally introduced in Ref.~\cite{wangEntanglementNonlocalityInfinite2017} by one of the present authors. The notation $mXY$ denotes that each party has $X$ measurement inputs, each input produces $Y$ possible outcomes, and the largest interaction range is $m$.
A 232-type TI contextuality witness takes the form of \begin{equation}\label{232-ineq}
\langle\mathcal{E}\rangle_{\text{TI}}=\Bigl\langle\sum_{x=0}^{2}J_x\sigma^{(1)}_x
+\sum_{x,y=0}^2J_{xy}\sigma^{(1)}_x\sigma^{(2)}_y\Bigl\rangle_{\text{TI}}
   \geq \mathcal{L},
\end{equation}
and a 322-type TI contextuality witness takes the form of 
\begin{equation}
    \begin{aligned}
\langle\mathcal{E}\rangle_{\text{TI}}=\Bigl\langle & \sum_{x=0}^{1}J_x\sigma^{(1)}_x
+\sum_{x,y=0}^1J_{xy}\sigma^{(1)}_x\sigma^{(2)}_y 
\\
+ &\sum_{x,z=0}^1J_{xz}\sigma^{(1)}_x\sigma^{(3)}_z\Bigl\rangle_{\text{TI}}
   \geq \mathcal{L},
\end{aligned}
\end{equation}
where $\sigma_x$ ($\sigma_y,\sigma_z$) can be arbitrary Hermitian matrices with eigenvalues $\pm1$ (not necessarily Pauli operators), and the expectation $\langle\mathcal{E}\rangle_{\text{TI}}$ is evaluated over TI reduced states. By TI reduced state, we mean that the state is a $m$-body reduced density matrix of a 1D infinite TI state. If there exist TI reduced states and observables such that the expectation value $\langle\mathcal{E}\rangle_{\text{TI}}$ is lower than the classical bound $\mathcal{L}$, the contextuality witness is said to be violated, and the corresponding quantum system is contextual—meaning that no local hidden-variable (LHV) model can account for the observed statistics \cite{bellEinsteinPodolskyRosen1964,nicolasbrunnerBellNonlocality2014,kochen1967problem,cabelloBellNonlocalityKochenSpecker2021,budroniKochenSpeckerContextuality2022}. When the violation attains its maximum over all possible quantum systems, the contextuality witness is said to be maximally violated; this maximum value is referred to as the quantum bound or quantum limit. We sometimes refer to a contextuality witness simply as an inequality throughout this paper, as no confusion is expected to arise.

To find violations of a TI contextuality witness, one needs to optimize the TI state and observables to break the classical bound. For convenience, we assume that the contextuality witness takes the form of Eq.~\eqref{232-ineq}. For fixed observables, optimizing the state is equivalent to computing the ground state of the TI Hamiltonian \begin{equation}\label{ti-ha}
H=\sum_{i=1}^{\infty}(\sum_{x=0}^{2} J_{x}\sigma^{(i)}_x+\sum_{x,y=0}^{2} J_{xy}\sigma^{(i)}_x\sigma^{(i+1)}_y),
\end{equation}
and then computing the two-body reduced state. Therefore, to obtain the optimal observables and state, it suffices to optimize the set of observables $\{\sigma_0,\sigma_1,\sigma_2\}$ such that the ground-state energy density of Eq.~\eqref{ti-ha} is lower than $\mathcal{L}$.

\begin{figure}[htbp]
\centering
\includegraphics[width=7.5cm]{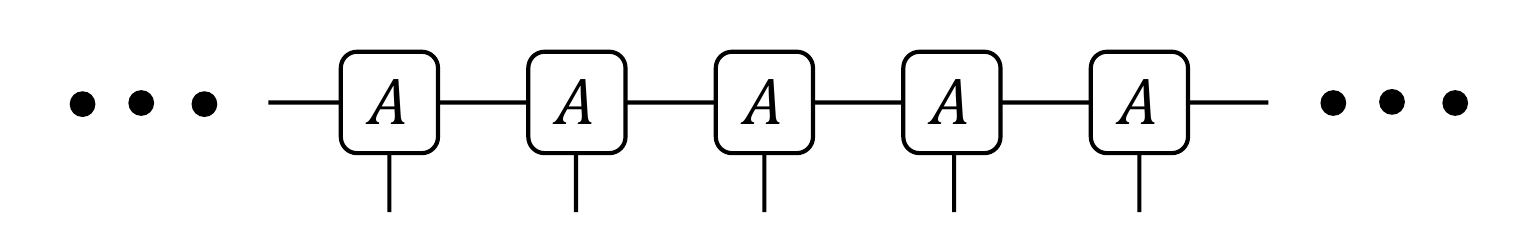}
\caption{A diagram representation of uMPS.}
\label{umps}
\end{figure}

To represent the ground state of an infinite 1D TI Hamiltonian, we use the uniform matrix product state (uMPS)~\cite{Valentine2018VUMPS}. A uMPS specified by a single tensor $A$ on an infinite 1D TI chain is defined as
\begin{equation}
    \ket{\Psi(A)} = \sum_{s_i} v_L^{\dagger}\left( \prod_{i \in \mathbb{Z}} A^{s_i}\right) v_R \ket{s},
\end{equation}
where $v_L, v_R$ are the left and right boundary vectors, and $A$ consists of $d$ matrices of size $D \times D$. Here, $d$ is the physical dimension, and $D$ is the bond dimension. A diagram representation is shown in Figure~\ref{umps}. 
Efficient algorithms, such as the time-dependent variational principle~\cite{haegemanTimeDependentVariationalPrinciple2011} and variational uMPS~\cite{Valentine2018VUMPS}, have been proposed to compute the ground state of an infinite 1D TI local Hamiltonian.

\subsection{Translation-symmetry breaking and periodic structure from MPS}
A key property of MPS is injectivity \cite{MPS-representation}, which we now introduce. The MPS space of $l$ sites spanned by $A$, a collection of $d$ matrices $A^s \in \mathbb{C}^{D \times D}$, is
\begin{equation}
    S_l = \{\sum_{s_1, \dots, s_l} \text{tr}(A^{s_1} \cdots A^{s_l} X)\ket{s_1 \cdots s_l} : X \in \mathbb{C}^{D \times D} \}.
\end{equation}
The MPS is said to be injective if $S_l$ reaches full rank for some $l$. This is equivalent to the condition that the transfer matrix
\begin{equation}
    E_A=\sum_s \bar{A}^s \otimes A^s
\end{equation}
has a single eigenvalue of magnitude 1. In the case relevant to this work, the transfer matrix 
may instead have $p>1$ eigenvalues of magnitude one forming the set
\begin{equation}
\left\{e^{2\pi i k/p}:k=0,\ldots,p-1\right\}.
\end{equation}
This peripheral spectrum implies a decomposition into $p$ extremal sectors related by one-site translations. Each extremal sector is invariant under translations by $p$ sites, but not under one-site translations. We therefore identify $p$ as the spatial period of the corresponding symmetry-broken ground-state sector. In this way, the period can be directly diagnosed from the peripheral spectrum of the MPS transfer matrix.

\begin{figure}[htbp]
\centering
\includegraphics[width=7cm]{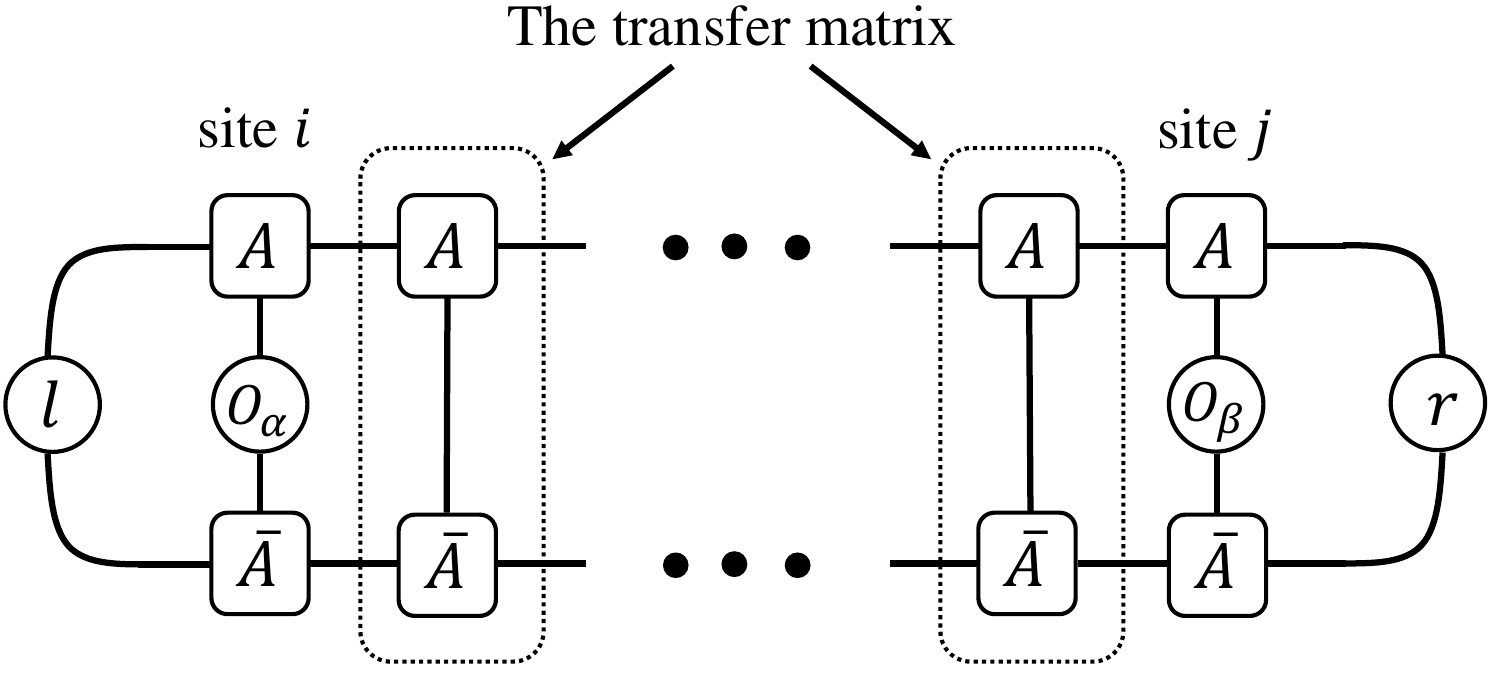}
\caption{The diagram representation of the correlation function $\langle O^i_{\alpha} O^j_{\beta}\rangle$. $l,r$ are the left and right fixed points of the transfer matrix $E_A$, respectively. $\langle O^i_{\alpha} O^j_{\beta}\rangle$ is governed by the products of $E_A$, and the behavior with respect to the distance $|j-i|$ is determined by the eigenvalues of $E_A$. If $E_A$ has only one eigenvalue of magnitude one, $\langle O^i_{\alpha} O^j_{\beta}\rangle$ decays exponentially with $|j-i|$. If $E_A$ has $p$ eigenvalues of magnitude one, $\langle O^i_{\alpha} O^j_{\beta}\rangle$ can contain persistent periodic 
components with period $p$.}
\label{correlationfunction}
\end{figure}

The two-body correlation function $\langle O^i_{\alpha} O^j_{\beta}\rangle$, with $O^i_{\alpha}$ ($O^j_{\beta}$) acting on site $i$ ($j$), defined as \begin{equation}
    \langle O^i_{\alpha} O^j_{\beta}\rangle=\bra{\Psi(A)}O^i_{\alpha} O^j_{\beta}\ket{\Psi(A)},
\end{equation}
has a diagrammatic representation, as shown in Figure \ref{correlationfunction}.

The case of $p>1$ corresponds to a translation symmetry-broken phase, where the correlation function does not decay exponentially but oscillates periodically with period $p$. If all the nonzero eigenvalues of $E_A$ lie on the unit circle and form $p$-th roots of unity, we say the state is strictly periodic, as correlation functions contain only exactly periodic components and no decaying terms.

\section{Contextuality as a diagnostic of translation-symmetry breaking in 1D TI systems}\label{sec:main-results}
Let $\langle\mathcal{E}\rangle_{\TI}\geq \mathcal{L}$ be a 232- or 322-type TI contextuality witness, and let
$H(\theta)$ be a one-parameter family of TI Hamiltonians that continuously interpolates between a
classical-bound Hamiltonian $H(\theta_0)$ and a quantum-optimal Hamiltonian $H(\theta_1)$, where
$H(\theta_0)$ saturates the classical bound and $H(\theta_1)$ attains the maximal quantum violation.
Our first main result, established analytically in two representative instances and supported numerically more broadly, is that at the maximal violation point $\theta=\theta_1$,
the ground states are \emph{strictly} $p_{\mathcal{Q}}$-periodic. In contrast, at
$\theta=\theta_0$ the ground-state manifold is highly degenerate and contains ground states with
multiple distinct periods; in particular, the smallest period satisfies $p_{\mathcal{L}}<p_{\mathcal{Q}}$.
In the interpolating families studied here, as $\theta$ is tuned away from $\theta_0$ into the contextual regime, this multiplicity of classical
periods is lifted and the system selects a single periodic structure (typically of period
$p_{\mathcal{Q}}$). Consequently, along such interpolations, our analytic and numerical evidence shows that entering the contextual regime coincides with selection of the $p_{\mathcal{Q}}$-periodic sector.

The degeneracy mechanism at the classical bound and its relation to multiple coexisting periods are
analyzed in Section~\ref{sec:classical}. Our second main result is that, for the models studied here,
the quantum-selected period also gives a compact finite-ring
benchmark. By applying TI-adapted variants of the Navascués-Pironio-Acín (NPA) hierarchy \cite{navascuesBoundingSetQuantum2007, navascuesConvergentHierarchySemidefinite2008}—including LTI and
periodic-boundary-condition (PBC) formulations—we show that the maximal quantum violation can
be saturated by periodic states and, moreover, that the TI witness is equivalent to a finite PBC
contextuality witness. Concretely, suppose the maximal violation is attained by a ground state with
period $p_{\mathcal{Q}}>1$, and that the same quantum limit is reproduced by the PBC-NPA with
$p_{\mathcal{Q}}$ parties. Then the induced $p_{\mathcal{Q}}$-partite PBC witness
\begin{equation}
\langle\mathcal{E}_{\PBC}\rangle
=\Bigl\langle\frac{1}{p_{\mathcal{Q}}}\sum_{i=1}^{p_{\mathcal{Q}}} T^{i}(\mathcal{E})\Bigr\rangle
\geq \mathcal{L},
\end{equation}
where $T$ denotes the one-site translation operator, is a valid contextuality witness. Furthermore,
$\langle \mathcal{E}_{\PBC} \rangle$ shares the same classical bound and quantum limit as its TI
counterpart $\langle\mathcal{E}\rangle_{\TI}$. The PBC reduction is discussed in
Section~\ref{sec:npa}.

This TI-to-PBC reduction is particularly significant operationally: it yields compact,
hardware-testable contextuality benchmarks on small rings. Simulations and experiments on such finite systems nevertheless capture key features of the ground-state physics in the thermodynamic limit. Practical benchmarking implications are discussed in
Section~\ref{sec:benchmark}.

The remainder of this section elaborates on translation-symmetry breaking in two models with exact ground states, corresponding to the 232- and 322-type TI contextuality witnesses.

\subsection{Exact example: a 232-type witness with strict period $p=3$} \label{sec2-232}
Consider a 232-type inequality in the form of \begin{equation}\label{ineq232-1}
   \Bigl\langle \sigma^{(1)}_1\sigma^{(2)}_0+\sigma^{(1)}_1\sigma^{(2)}_1-\sigma^{(1)}_2\sigma^{(2)}_0+\sigma^{(1)}_2\sigma^{(2)}_1 \Bigl\rangle_{\text{TI}} \geq -2,
\end{equation}
and the Hamiltonian reads 
\begin{equation}\label{232ha}
    \begin{aligned}
    H=\sum_{i=1}^{\infty} &\sigma^{(i)}_1\sigma^{(i+1)}_0+\sigma^{(i)}_1\sigma^{(i+1)}_1
    \\
    -&\sigma^{(i)}_2\sigma^{(i+1)}_0+\sigma^{(i)}_2\sigma^{(i+1)}_1.
\end{aligned}
\end{equation}
The maximal violation is achieved when $d=5$. 

We have derived an analytical solution to the maximal violation problem of this inequality (see Appendix~\ref{appendix:232analytical} for details). The optimal values of $\sigma_0,\sigma_1,\sigma_2$ are \begin{equation}\label{232ob-1}{\footnotesize
\begin{aligned}
    &\sigma_0 =     \begin{pmatrix*}[r]
        1 & 0 & 0 & 0 & 0\\
        0 & 1 & 0 & 0 & 0\\
        0 & 0 & -1 & 0 & 0\\
        0 & 0 & 0 & 1 & 0\\
        0 & 0 & 0 & 0 & 1
    \end{pmatrix*},   \sigma_1 = -     \begin{pmatrix*}[r]
        0 & 0 & 1 & 0 & 0\\
        0 & 1 & 0 & 0 & 0\\
        1 & 0 & 0 & 0 & 0\\
        0 & 0 & 0 & 0 & 1\\
        0 & 0 & 0 & 1 & 0
    \end{pmatrix*},\\
       &\sigma_2 =     \begin{pmatrix*}[r]
        1 & 0 & 0 & 0 & 0\\
        0 & 1 & 0 & 0 & 0\\
        0 & 0 & 1 & 0 & 0\\
        0 & 0 & 0 & 1 & 0\\
        0 & 0 & 0 & 0 & -1
    \end{pmatrix*}.
\end{aligned}}
\end{equation}

The ground state is given by the uMPS $\ket{\Psi(A)}$ with A being \begin{equation}\small \label{232mps1}{\footnotesize
    \begin{aligned}
    & A^0 = \begin{pmatrix*}[r]
         0&        0&       0&  0\\
         0&        0&       0&  0\\
         \alpha&  -\alpha&  0&  0\\
         \beta&   -\beta&   0&  0
    \end{pmatrix*},
     A^1 = \begin{pmatrix*}[r]
          1/2&   1/2&  0&  0\\
         -1/2&  -1/2&  0&  0\\
          0&   0&  0&  0\\
          0&   0&  0&  0
    \end{pmatrix*},\\
    & A^2 = \begin{pmatrix*}[r]
          0&        0 &      0&  0\\
          0   &     0  &     0&  0\\
         -\beta&    \beta&   0&  0\\
          \alpha&  -\alpha&  0&  0
    \end{pmatrix*} ,
     A^3 = \begin{pmatrix*}[r]
         0&  0&  0&  1/2\\
         0&  0&  0&  1/2\\
         0&  0&  0&  0\\
         0&  0&  0&  0
    \end{pmatrix*} ,
    \\
     &A^4 = \begin{pmatrix*}[r]
         0&  0&  1/2&  0\\
         0&  0&  1/2&  0\\
         0&  0&  0&  0\\
         0&  0&  0&  0
    \end{pmatrix*}, \\ 
\end{aligned}}
\end{equation}
where $\alpha=\sqrt{\frac{2-\sqrt{2}}{8}},\beta=\sqrt{\frac{2+\sqrt{2}}{8}}$.
The ground-state energy density is $-\frac{2}{3}(2+\sqrt{2})$. The fact that Eq.~\eqref{232mps1} is the exact ground state of Eq.~\eqref{232ha}, with Eq.~\eqref{232ob-1} as the corresponding observable, follows from the agreement between its energy density and the NPA value of Eq.~\eqref{ineq232-1} (see Appendix~\ref{appendix-npa232}). We therefore obtain an analytical solution for both the ground state of the Hamiltonian and the quantum limit.

\begin{figure}[htbp]
\centering
\includegraphics[width=8cm]{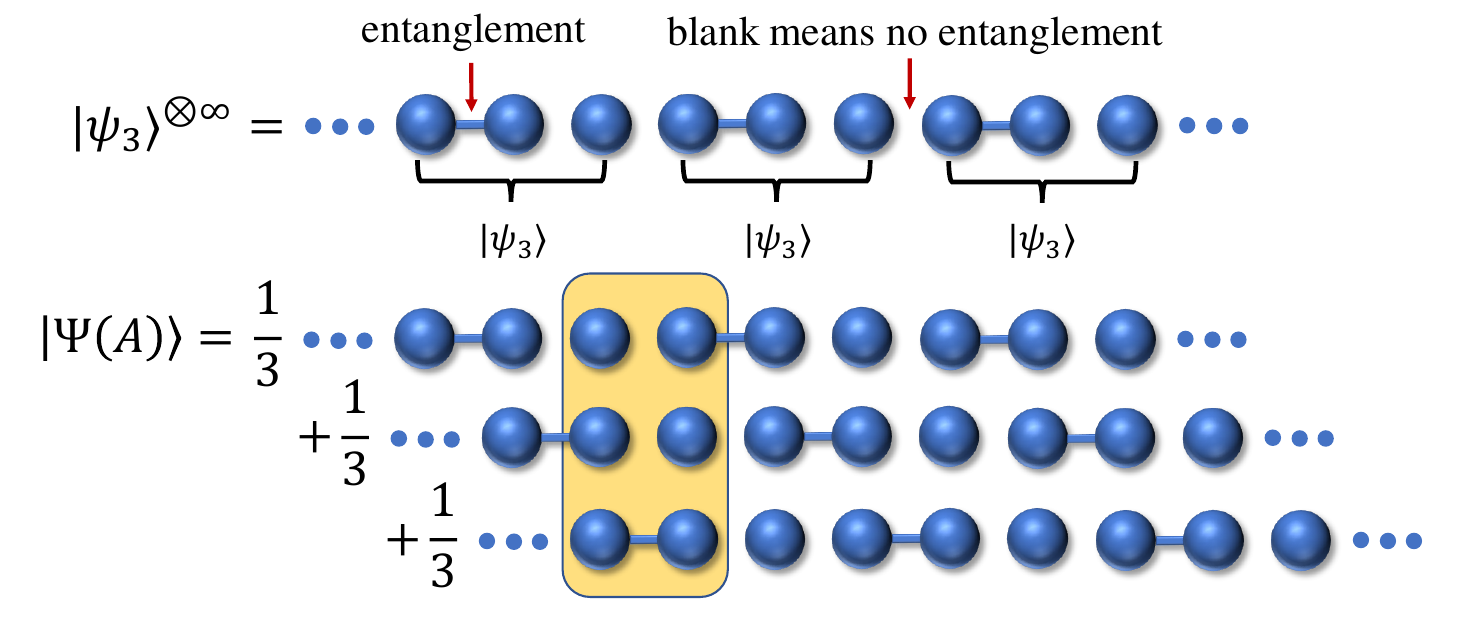}
\caption{Structure of the ground state $\ket{\Psi(A)}$. $\ket{\Psi(A)}$ is the symmetrization of the state $\ket{\psi_3}^{\otimes \infty}$, and $\ket{\psi_3}$ is a state that is a product of a two-partite entangled state and a single-partite state. The yellow box shows the symmetrized two-site reduced density matrix.}
\label{232-state-structure}
\end{figure}

Using Eq.~\eqref{232mps1}, one can compute the nonzero eigenvalues of the transfer matrix exactly; they are $e^{2\pi i k / 3}$ for $k = 0, 1, 2$, implying $p_{\mathcal{Q}} = 3$. In fact, the uMPS $\ket{\Psi(A)}$ is block injective and has a decomposition \begin{equation}
    \ket{\Psi(A)}=\frac{1}{\sqrt{3}}\sum_{i=0}^2 T^i(\ket{\psi_3}^{\otimes \infty}),
\end{equation} where $\ket{\psi_3}$ is a state defined on three sites, i.e., \begin{equation}
    \ket{\psi_3}=(-\beta\ket{30}-\alpha\ket{32}-\alpha\ket{40}+\beta\ket{42})\otimes\ket{1},
\end{equation} and $T$ is the one-site translation operator. Figure~\ref{232-state-structure} shows the structure of the ground state.

The classical bound $-2$ can be saturated by using \begin{equation}\label{232-cb}{\footnotesize
    \begin{aligned}
        &\sigma_0=\begin{pmatrix*}[r]
        1 & 0 & 0 & 0 & 0\\
        0 & 1 & 0 & 0 & 0\\
        0 & 0 & -1 & 0 & 0\\
        0 & 0 & 0 & 1 & 0\\
        0 & 0 & 0 & 0 & 1
    \end{pmatrix*},    \sigma_1=-\begin{pmatrix*}[r]
        1 & 0 & 0 & 0 & 0\\
        0 & 1 & 0 & 0 & 0\\
        0 & 0 & -1 & 0 & 0\\
        0 & 0 & 0 & 1 & 0\\
        0 & 0 & 0 & 0 & -1
    \end{pmatrix*},\\   & \sigma_2=\begin{pmatrix*}[r]
        1 & 0 & 0 & 0 & 0\\
        0 & 1 & 0 & 0 & 0\\
        0 & 0 & 1 & 0 & 0\\
        0 & 0 & 0 & 1 & 0\\
        0 & 0 & 0 & 0 & -1
    \end{pmatrix*},
    \end{aligned}}
\end{equation} 
and the ground space is highly degenerate and contains ground states of periods ranging from 1 to 8 (see Section \ref{sec:classical} for the determination of the classical periods). The smallest period is $P_\mathcal{L}=1$. As we interpolate between Eq.~\eqref{232ob-1} and Eq.~\eqref{232-cb} by applying a rotation $R(\theta)$ of $\sigma_1$ in the subspace spanned by $(\ket{0},\ket{2})$ and the subspace spanned by $(\ket{3},\ket{4})$, spontaneous symmetry breaking occurs. The nonzero eigenvalues of the transfer matrix of Eq.~\eqref{232mps1} are exactly $\{e^{2\pi ik/3}: k=0,1,2\}$, implying $p_\mathcal{Q}=3$. Remarkably, when $\theta\in(0,\pi/4]$, the corresponding ground states computed by the time-dependent variational principle (TDVP) \cite{haegemanTimeDependentVariationalPrinciple2011} algorithm all exhibit a period of 3 to numerical precision $10^{-8}$. The classical bound $-2$ marks the symmetry-change point.

The finite PBC inequality 
\begin{equation}\label{232-pbc}
    \begin{aligned}
\frac{1}{3}\sum_{i=1}^3\Bigl\langle &\sigma^{(i)}_1\sigma^{(i+1)}_0+\sigma^{(i)}_1\sigma^{(i+1)}_1-
\\
& \sigma^{(i)}_2\sigma^{(i+1)}_0+\sigma^{(i)}_2\sigma^{(i+1)}_1 \Bigl\rangle \geq -2
\end{aligned}
\end{equation}
is a contextuality witness, and the quantum limit is also $-\frac{2}{3} \left(2+\sqrt{2}\right)$.

The 232-type inequality Eq.~\eqref{ineq232-1} has additional intriguing properties. First of all, without TI, the finite two-partite inequality \begin{equation}\label{ineq232:2party}
   \langle \sigma^{(1)}_1\sigma^{(2)}_0+\sigma^{(1)}_1\sigma^{(2)}_1-\sigma^{(1)}_2\sigma^{(2)}_0+\sigma^{(1)}_2\sigma^{(2)}_1 \rangle \geq -2
\end{equation}
is also a tight inequality. In fact, Eq.~\eqref{ineq232:2party} is a 3322-scenario inequality equivalent to the CHSH inequality \cite{collinsRelevantTwoQubit2004}, and the maximal violation is $-2\sqrt{2}$. The $n$-partite inequality with open boundary conditions (OBC) given by \begin{equation}\label{232:nparty}
\begin{aligned}
\frac{1}{n-1}\sum_{i=1}^{n-1}\Bigl\langle &\sigma^{(i)}_1\sigma^{(i+1)}_0+\sigma^{(i)}_1\sigma^{(i+1)}_1-\\
&
\sigma^{(i)}_2\sigma^{(i+1)}_0+\sigma^{(i)}_2\sigma^{(i+1)}_1 \Bigl\rangle \geq -2 
\end{aligned}
\end{equation} is also a valid inequality for any $n>2$, since the classical bound can always be saturated by an LHV model (see details in Section \ref{subsec:lhv}) given by \begin{equation}
Q(a_{1,0},a_{1,1},a_{1,2},\ldots,a_{n,0},a_{n,1},a_{n,2}) = 1,
\end{equation} 
where, for $j=1,\ldots,n$, $(a_{j,0},a_{j,1},a_{j,2})=(0,0,0)$ for odd $j$, and $(a_{j,0},a_{j,1},a_{j,2})=(0,1,0)$ for even $j$.

\begin{figure}[htbp]
\centering
\includegraphics[width=8.2cm]{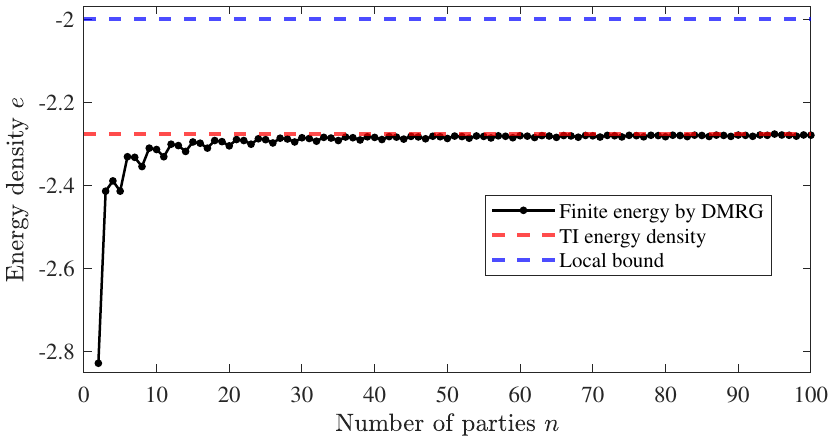}
\caption{The energy densities of Eq.~\eqref{232:nparty} (left-hand side with observables Eq.~\eqref{232ob-1}) computed using density matrix renormalization group (DMRG) with OBC \cite{whiteDensitymatrixAlgorithmsQuantum1993}. Note that DMRG sometimes falls into local minima for this model, and the final energy seems to depend on the initial tensor. Each point shows the lowest energy found over 100 runs of DMRG for each $n\in \{3,\ldots,100\}$. }
\label{fig232finitesizescaling}
\end{figure}

As in the TI case, the left-hand side of Eq.~\eqref{232:nparty} defines a Hamiltonian $H_{\OBC}$ with observables yet to be specified. Remarkably, for all system sizes $n$ up to $100$, choosing the observables according to Eq.~\eqref{232ob-1} leads to a ground-state energy of $H_{\OBC}$ that violates Eq.~\eqref{232:nparty}, as shown in Figure~\ref{fig232finitesizescaling}. Moreover, for $n=2$ and $n=3$, Eq.~\eqref{232ob-1} achieves the maximal quantum violation, as certified by the NPA hierarchy. We conjecture that Eq.~\eqref{232ob-1} attains the maximal quantum violation for all $n$. 

Figure~\ref{fig232finitesizescaling} provides a direct visualization of the finite-size scaling under OBC. The finite-size energy densities converge to the TI energy density as $n$ goes to infinity. Examining nearest-neighbor two-body energies of the ground state reveals a clear pattern: when $n = 3k$ or $n = 3k + 2$, the two-body energies exhibit a periodic structure composed of $\{-2\sqrt{2}, -2, -2\}$, except for the last one or two terms. In contrast, when $n = 3k + 1$, the period is disrupted by edge effects. In the thermodynamic limit, the energy pattern becomes perfectly periodic as $\{-2\sqrt{2}, -2, -2\}$, recovering the exact TI energy density.

\subsection{Exact example: a 322-type witness with strict period $p=4$}
\label{sec3}
This subsection presents details for the 322-type models that exhibit contextuality and translation-symmetry breaking. We consider a 322-type inequality in the form of 
\begin{equation}
    \begin{aligned}
        \langle&-3\sigma_0^{(1)} + \sigma_1^{(1)} + \sigma_0^{(1)}\sigma_0^{(2)}+\sigma_0^{(1)}\sigma_1^{(2)}+\sigma_1^{(1)}\sigma_0^{(2)}
        \\ &
        -\sigma_1^{(1)}\sigma_1^{(2)}+\sigma_0^{(1)}\sigma_0^{(3)}-\sigma_1^{(1)}\sigma_0^{(3)}+\sigma_1^{(1)}\sigma_1^{(3)}\rangle \geq -3.
\end{aligned}
\end{equation}
$\sigma_0,\sigma_1$ are parameterized as \begin{align}
	&\sigma_0 = \begin{pmatrix}
		1 &  0 & 0\\ 
		0 & -1 & 0\\
        0 &  0 & 1
	\end{pmatrix},
	\\&\sigma_1 = 
	\begin{pmatrix}
		\cos(2\theta) & -\sin(2\theta)  &0 \\ 
	   -\sin(2\theta) & -\cos(2\theta)  &0 \\
               0 &         0  &-1
	\end{pmatrix}.
    \label{322 parameterized observables}
\end{align}
The Hamiltonian $H(\theta)$ given by \begin{equation}
    \begin{aligned}
    H=\sum_{i=1}^{\infty} &-3\sigma_0^{(i)} + \sigma_1^{(i)} +\sigma_0^{(i)}\sigma_0^{(i+1)}+\sigma_0^{(i)}\sigma_1^{(i+1)}\\&
    +\sigma_1^{(i)}\sigma_0^{(i+1)}-\sigma_1^{(i)}\sigma_1^{(i+1)}+\sigma_0^{(i)}\sigma_0^{(i+2)}\\&-\sigma_1^{(i)}\sigma_0^{(i+2)}+\sigma_1^{(i)}\sigma_1^{(i+2)}
\end{aligned}
\end{equation}
depends on a single parameter $\theta$. We use the TDVP algorithm to compute the ground state $\ket{\Psi(A_{\theta})}$ and energy density $e_{\theta}$ of $H(\theta)$. When $\theta=0$, we have $\sigma_1=\sigma_0$, $e_0 = -3$, and $H(0)$ becomes diagonal. In fact, the ground space of $H(0)$ is highly degenerate, and it contains ground states of various distinct periods consisting of $\{1,3,4,5,6,7,8,9,10\}$. The smallest period of the ground space is $p_{\mathcal{L}}=1$. 

When $\theta=0.78524585$, the ground-state energy density reaches the quantum limit \cite{Yang2022}, and the bond dimension $D$ is 5. Using the technique developed in Section~\ref{subsubsec-322-exact}, we can derive the exact solution for $\theta$, which is $\theta=\pi/4$, and the exact energy density is $-(5+\sqrt{2})/2$. The nonzero eigenvalues of the transfer matrix $E_{A_{\theta}}$ are $\{1,-1,i,-i\}$, which means that the period is $p_{\mathcal{Q}}=4$. This means that the one-site translation symmetry is broken in order to achieve the quantum limit. For sampled values of $\theta \in (0,\pi/4]$, the uMPS ground states exhibit the peripheral transfer spectrum $\{1,-1,i,-i\}$ up to a numerical precision of $10^{-8}$, supporting period-$4$ selection throughout the contextual region of this interpolation.

For example, already at \(\theta=0.05\), where the violation is small
(\(e_\theta=-3.0024855\)), the peripheral transfer spectrum still
contains \(\{1,-1,i,-i\}\) to numerical precision. This supports the
interpretation that, along this interpolation, the classical-bound point
marks the onset of period-4 sector selection.

The other 322-type inequalities listed in Table~\ref{tab:322} show the same period-selection pattern to numerical precision $10^{-8}$. The quantum values \(Q\) in Table~\ref{tab:322} are taken from Ref.~\cite{Yang2022}, where the corresponding translation-invariant contextuality witnesses were optimized. In the present work, we use these values as reference quantum optima and focus on the ground-state structure of the Hamiltonians that realize them. For each row, the quantum-selected period $p_{\mathcal{Q}}$ is extracted from the peripheral spectrum of the optimized uMPS transfer matrix: when the unit-modulus eigenvalues form the set \(\{e^{2\pi i k/p_{\mathcal{Q}}}\}_{k=0}^{p_{\mathcal{Q}}-1}\), we identify the corresponding ground-state sector as \(p_{\mathcal{Q}}\)-periodic. The integer \(D\) denotes the bond dimension of the uMPS representative used for that optimal ground state. The classical period \(p_L\) is the smallest period appearing in the classical-bound ground-sector construction described in Section~\ref{sec:classical}, where the relevant sectors are obtained from the corresponding LHV/domino-loop representation.

\begin{table*}[htbp]
    \caption{The periods associated with the ten 322-type inequalities at maximal quantum violation. The $J$'s are the coefficients of the inequalities. $d_0$ is the physical dimension at which each model reaches the maximal quantum violation. $\mathcal{L}$ is the classical bound. $\mathcal{Q}$ is the quantum limit. $D$ is the bond dimension of the ground states that reach the quantum limit. $p_{\mathcal{L}}$ is the smallest period among the ground states that reach the classical bound, and $p_{\mathcal{Q}}$ is the period of the ground state that achieves the quantum limit.}
    \centering
    \renewcommand\arraystretch{1.1}
    \setlength{\tabcolsep}{2.2mm}{
    \begin{tabular}{|c|c|c|c|c|c|c|c|c|c|c|c|c|c|c|c|c|c|c|c|c|}	
    \hline
No. & $J_0$ &$J_1$ & $J_{00}^{12}$ & $J_{01}^{12}$ & $J_{10}^{12}$ & $J_{11}^{12}$ & $J_{00}^{13}$ & $J_{01}^{13}$ & $J_{10}^{13}$ & $J_{11}^{13}$ & $d_0$ & $\cal{L}$ & ${\cal{Q}}$ & $D$ & $p_{\cal{L}}$ & $p_{\cal{Q}}$
\\ \hline						
    1 &	 -6&  0&  2 &	3 &	3 &	-2 &	3 &	-1 &	-1 &	1 & 3& -6 & -6.32747 &5 & 1 & 4	
    \\ \hline									 	 	 
    2&	-4 &	2 &	2 &	2 &	2 &	-4 &	1 &	-1 &	-1 &	3 & 3& -6 & -6.33711&5 & 1 & 4
    \\	\hline								 	 	
    3&	-3 &	1 &	1 &	1 &	1 &	-1 &	1 & 0 &	-1 & 1 & 3 & -3	& -3.20710&5  & 1 & 4
    \\ \hline
    4&	-2 &	-2 &	-2 &	1 &	-1 &	-2 &	1 &	0 &	2 &	1 & 3& -4	& -4.14623&5 & 1 & 4
    \\ \hline				 	
    5	& 	-11 &	1 &	5 &	2 &	2 &	-1& 	4 &	-1 &	-2 	&1 &3 &-8 &	-8.12123&5 & 3 & 4
    \\ \hline									 		
    6&	-3 &	-3 &	2 &	2 	&-1 &	2 &	1 &	1 &	-1 &	0 &	4 & -4 & -4.10309&6 &  2 &	5		
\\ \hline
7&		-3 &	-3& 	2 &	2 &	2 &	-3 &	1 &	0 &	-1 &	2 & 5 & -5	&-5.29851&6 & 2 & 5	
\\ \hline
8&	-2&	 	-4 &		-2 &		2 &		2 &		2 	&	1 &		0 &		0 &		1 &	5&-4& -4.33137&6 & 1 & 5	
\\ \hline
9	&			0 &		-4 &		2 &		2 &		-2 &		2 &		0 &		1 &		-1 &		0 	&	5& -4	& -4.41421&5 & 2 & 4	 \\ \hline	
10	&			-2 &		2 &		2 &		-2 &		-2 &		-4 &		1 &		1 &		1 &		2 	&	5 & -5	& -5.26969&6 	&	 2 & 5\\ \hline
    \end{tabular}}
    \label{tab:322}
\end{table*}
The 322- and 232-type inequalities discussed in this section are particularly well suited for quantum simulation and benchmarking of quantum devices. By simulating the ground states of the parameterized Hamiltonians, one can simultaneously observe symmetry breaking and manifestations of contextuality. Moreover, the finite versions of the TI inequalities enable direct simulation of finite systems.

\section{Why classical saturation is highly degenerate: TI classical extensions and domino loops}
\label{sec:classical}
This section shows that the ground-state structure corresponding to the classical bound can be explained through the LHV model and a domino-tiling construction, and we demonstrate how to construct Hamiltonians that saturate the classical bound, and explain the mechanism of the multiple periods of the classical ground states. We begin by introducing the local hidden-variable (LHV) model, and then elaborate on the connection between TI distributions and domino tiling, and finally show how to construct TI Hamiltonians to reproduce the classical bound.

\subsection{The LHV model of TI contextuality witnesses}\label{subsec:lhv}
Consider a Bell scenario with infinitely many parties arranged in a 1D chain labeled by integers. At each site $i\in\mathbb{Z}$, party $i$ performs a measurement $x_i\in\{0,\ldots,X-1\}$ yielding an outcome $a_i\in Y$. TI requires that the statistics of any $m$ consecutive parties coincide with those of parties $1,\ldots,m$, i.e.,
\begin{equation}
\begin{aligned}
    &P_{k,\ldots,m+k-1}(a_k,\ldots,a_{m+k-1}|x_k,\ldots,x_{m+k-1})\\
   =&P_{1,\ldots,m}(a_1,\ldots,a_m|x_1,\ldots,x_m),
\end{aligned}
\end{equation}
for all $k\in\mathbb{Z}$. Hence, a TI Bell scenario is fully specified by the integers $m$, $X$, and $|Y|$.

For an $m$-partite classical system with inputs $\{x_i\}$ and outputs $\{a_i\}$, the correlations $P_C(a_1,\ldots,a_m|x_1,\ldots,x_m)$ admit an LHV decomposition,
\begin{equation}\label{LHV}
   \begin{aligned}
&P_C(a_1,\ldots,a_m|x_1,\ldots,x_m)
   \\ = &\!\int\! P(\lambda)\prod_{i=1}^mP_{\lambda}(a_i|x_i)\,d\lambda,
\end{aligned} 
\end{equation}
where $\lambda$ denotes hidden variables. The distributions $\{P_{\lambda}(a_i|x_i)\}$ and $P(\lambda)$ constitute an LHV model for $P_C$. 
By Fine's theorem \cite{fineHiddenVariablesJoint1982}, the existence of an LHV model for $P_C$ is equivalent to the existence of a distribution $Q(\vec{a_1},\vec{a_2}, \ldots, \vec{a_m})$, with $\vec{a_i}=(a_{i,0},a_{i,1},\ldots,a_{i,X-1})$ such that \begin{equation}\label{fine theorem}
\begin{aligned}
    & P_C(a_1,\ldots,a_m \mid x_1,\ldots,x_m)
\\ & =
\sum_{\substack{
\vec a_1,\ldots,\vec a_m \\
a_{i,x_i}=a_i\ \forall i\in\{1,\ldots,m\}
}}
Q(\vec a_1,\ldots,\vec a_m).
\end{aligned}
\end{equation}
We also refer to $Q$ as the LHV model. 

For an $m$-partite quantum system, the correlations $P(a_1,\ldots,a_m|x_1,\ldots,x_m)$ take the form of \begin{equation}
P(a_1,\ldots,a_m|x_1,\ldots,x_m)
   =\text{tr}\!\left[\bigotimes_{i=1}^m E_{a_i|x_i}\,\rho\right],
\end{equation}
where $\rho$ is a quantum state defined on the $m$ systems, and $\{E_{a_i|x_i}\}$ are measurement operators satisfying $\sum_{a_i} E_{a_i|x_i}=\mathbb{I}$. Bell showed that there exist quantum correlations that cannot be explained by any LHV model, thereby rejecting the LHV description of quantum mechanics. In standard Bell tests, space-like separation enforces the independence required by~(\ref{LHV}); however, in many-body systems, this is generally impractical, and a violation of~(\ref{LHV}) is interpreted as contextuality~\cite{kochen1967problem,cabelloBellNonlocalityKochenSpecker2021,budroniKochenSpeckerContextuality2022}.

An $m$-partite classical distribution $P_C(a_1,\ldots,a_m|x_1,\ldots,x_m)$ admits a TI classical extension if there exists an infinite TI distribution $Q(\ldots,\vec{a}_{-1},\vec{a_{0}},\vec{a_{1}},\ldots)$ such that 
\begin{equation}\label{fine theorem}
\begin{aligned}
    & P_C(a_1,\ldots,a_m | x_1,\ldots,x_m)
 \\= &    
\sum_{\substack{
\vec a_1,\ldots,\vec a_m \\
a_{i,x_i}=a_i\ \forall i\in\{1,\ldots,m\}
}}
Q_m(\vec a_1,\ldots,\vec a_m),
\end{aligned}
\end{equation}
where $Q_m$ is the marginal distribution of $Q$ at sites $1,\ldots,m$.

A distribution $P_{1,\ldots,m}(a_1,\ldots,a_m|x_1,\ldots,x_m)$ admits a TI quantum extension if there exist a local Hilbert space $\mathcal{H}$, measurement operators $\{E_{a|x}\}$ satisfying $\sum_a E_{a|x}=\mathbb{I}$, and a TI quantum state $\rho$ on the infinite chain such that
\begin{equation}
\begin{aligned}
    & P_{1,\ldots,m}(a_{1},\ldots,a_m|x_1,\ldots,x_m)
   \\ = & \text{tr}\!\left[\bigotimes_{j=1}^m E_{a_j|x_j}\,\rho_{1,\ldots,m}\right],
\end{aligned}
\end{equation}
where $\rho_{1,\ldots,m}$ is the reduced state of $\rho$ on sites $1,\ldots,m$.

As shown in Ref.~\cite{wangEntanglementNonlocalityInfinite2017}, the set of $m$-nearest-neighbor correlations admitting a TI classical extension forms a polytope defined by finitely many linear inequalities (facets). 
For dichotomic outcomes ($Y=\{0,1\}$) and two measurements per site ($X=2$), if $m=3$ (the distribution is $P(abc|xyz)$), the facet (contextuality witness) takes the form of \begin{equation}\label{322-expectation}
\begin{aligned}
    & \sum_{x\in\{0,1\}}J_xE_x
+\sum_{x,y\in\{0,1\}}J_{xy}E_{xy}
\\+&\sum_{x,z\in\{0,1\}}J_{xz}E_{xz}
   \geq \mathcal{L},
\end{aligned}
\end{equation}
where \begin{equation*}
    \begin{aligned}
        &E_x^{(1)}=\sum_{a}(-1)^{a}P(a|x), \\
        &E_{xy}=\sum_{a,b}(-1)^{a+b}P(ab|xy),\\
        &E_{xz}=\sum_{a,c}(-1)^{a+c}P(ac|xz).
    \end{aligned}
\end{equation*}

For a TI quantum system, each observable $x$ at site $i$ can be represented by an operator $\sigma^i_x$ with eigenvalues $\pm1$, and the contextuality witness reads
\begin{equation}
\begin{aligned}
    \langle\mathcal{E}\rangle_{\text{TI}}=&\Bigl\langle\sum_{x \in \{0,1\}} J_x\sigma^{(1)}_x
+\sum_{x,y \in \{0,1\}}J_{xy}\sigma^{(1)}_x\sigma^{(2)}_y\\+&\sum_{x,z \in \{0,1\}}J_{xz}\sigma^{(1)}_x\sigma^{(3)}_z\Bigl\rangle_{\text{TI}}
   \geq \mathcal{L},
\end{aligned}
   \label{nnn-ineq}
\end{equation}
referred to as a 322-type TI contextuality witness. Violation of Eq.~\eqref{nnn-ineq} by the one- and two-body correlators of a TI system demonstrates the impossibility of a classical (LHV) description.

\subsection{TI LHV models and domino tiles}
Note that Fine's theorem states that any classical correlation $P(a_1a_2a_3|x_1x_2x_3)$ (the 322-scenario) can be written as a convex combination of local deterministic (LD) distributions \cite{fineHiddenVariablesJoint1982,ScaraniBook,nicolasbrunnerBellNonlocality2014}. $P$ is LD if it can be decomposed as $$P(a_1a_2a_3|x_1x_2x_3)=\delta_{a_1=f_1(x_1)}\delta_{a_2=f_2(x_2)}\delta_{a_3=f_3(x_3)},$$ where $f_1,f_2,f_3$ are deterministic functions. Formally, $P(a_1a_2a_3|x_1x_2x_3)$ is classical if there exists a probability distribution $Q(a_{1,0},a_{1,1},a_{2,0},a_{2,1},a_{3,0},a_{3,1})$ such that \begin{equation}\label{finetheorem}
\begin{split}
&P(a_1a_2a_3 | x_1x_2x_3)
\\=
&\sum_{\substack{
a_{1,x_1}=a_1\\
a_{2,x_2}=a_2\\
a_{3,x_3}=a_3
}}
Q(a_{1,0},a_{1,1},a_{2,0},a_{2,1},a_{3,0},a_{3,1}).
\end{split}
\end{equation}
Here, each component of $Q$ is the probability of the corresponding LD distribution. For example, $Q(a_{1,0}=1,a_{1,1}=0,a_{2,0}=1,a_{2,1}=0,a_{3,0}=1,a_{3,1}=0)$ gives the probability of $P$ taking the LD distribution $\delta(a_1,1-x_1)\delta(a_2,1-x_2)\delta(a_3,1-x_3)$. 

For a distribution $Q$, the necessary and sufficient condition for $Q$ to have a TI extension is that $Q$ satisfy the local translationally invariant (LTI) condition \cite{wangEntanglementNonlocalityInfinite2017}, i.e.,
\begin{equation}
    \begin{aligned}
        &\sum_{a_{3,0},a_{3,1}}Q(a_{1,0},a_{1,1},a_{2,0},a_{2,1},a_{3,0},a_{3,1}) \\
        = &\sum_{a_{1,0},a_{1,1}}Q(a_{1,0},a_{1,1},a_{2,0},a_{2,1},a_{3,0},a_{3,1}).
    \end{aligned}
\end{equation}
The set of $Q$ satisfying the LTI condition constitutes a convex polytope, denoted by LTI-$Q$ polytope.
Therefore, for a classical distribution $P$ to be a valid TI marginal, one simply seeks an LTI $Q$ such that Eq.~\eqref{finetheorem} holds. The set of classical TI marginals $P$, referred to as TI-C, forms a convex polytope, as the projection from $Q$ to $P$ is linear. After projecting TI-C onto the set of correlators, we finally arrive at the convex polytope of correlators, whose facets take the form of Eq.~\eqref{nnn-ineq}.

\begin{figure}[htbp]
\centering
\includegraphics[width=8.2cm]{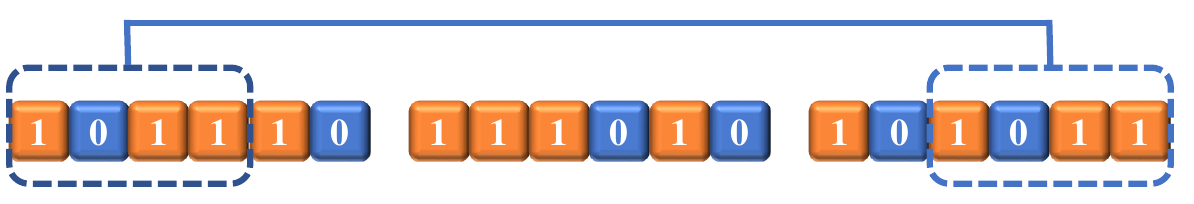}
\caption{An example of a domino loop. There are three domino tiles, where the last four numbers of each tile coincide with the first four numbers of the next tile. The last four numbers of the final tile match the first four numbers of the first tile, forming a closed loop. This domino loop is irreducible, as no pair among the three tiles can form a closed loop on its own.
}
\label{domino}
\end{figure}

Because the facet inequality can be determined by a set of correlator vertices $S_E$, we can identify a set of $Q$ vertices, denoted by $S_Q$, whose projections coincide with $S_E$. Then, the distributions $P$ generated by $S_Q$ all give the classical bound. The vertices of the LTI-$Q$ polytope can be classified into two groups; one is that the vertex is itself a TI LD distribution, and the other is that the vertex is a convex combination of LD distributions. The former case is trivial. We therefore consider the latter case and write $Q=\sum_{i=1}^k\lambda_iQ_i$, where each $Q_i$ is an LD distribution, meaning that there is only one configuration in $Q_i$ equal to 1 and all others equal 0. For each $Q_i$, we assume $Q_i(a_{1,0}=y_{i,1},a_{1,1}=y_{i,2},a_{2,0}=y_{i,3},a_{2,1}=y_{i,4},a_{3,0}=y_{i,5},a_{3,1}=y_{i,6})=1.$
Let $\vec{y_i}=(y_{i,1},y_{i,2},y_{i,3},y_{i,4},y_{i,5},y_{i,6})$ for $i=1,\ldots,k$.  
If after a suitable sorting of $\{\vec{y_i}\}$, the condition 
\begin{equation}
(y_{i,3},y_{i,4},y_{i,5},y_{i,6})=(y_{i+1,1},y_{i+1,2},y_{i+1,3},y_{i+1,4})
\end{equation} holds for any $i$, where $i+1\equiv1 \text{ if } i=k$, we say that $\{\vec{y_i}\}_{i=1}^k$ admits a domino loop, and each $\vec{y_i}$ is a domino tile. Figure~\ref{domino} is an example of a domino tiling. 
The domino loop is irreducible if no proper subset of $\{\vec{y_i}\}_{i=1}^k$ forms a valid domino loop.
It was proved in \cite{wangEntanglementNonlocalityInfinite2017} that all vertices of the LTI-$Q$ polytope admit irreducible domino loops.

\subsection{Constructing Hamiltonians achieving the classical bound} 
In the following, we show that the states saturating the classical bound are periodic, as revealed by the LHV model and a domino-tiling construction. For the 322-type inequality (first model in Table~\ref{tab:322}) in the form of 
\begin{equation}\label{322-model1}
    \begin{aligned}
    \langle &-6\sigma_0^{(1)} + 2\sigma_0^{(1)}\sigma_0^{(2)}+3\sigma_0^{(1)}\sigma_1^{(2)}+3\sigma_1^{(1)}\sigma_0^{(2)}\\&-2\sigma_1^{(1)}\sigma_1^{(2)}+3\sigma_0^{(1)}\sigma_0^{(3)}-\sigma_0^{(1)}\sigma_1^{(3)}-\sigma_1^{(1)}\sigma_0^{(3)}\\&+\sigma_1^{(1)}\sigma_1^{(3)}\rangle_{\text{TI}} \geq -6,
\end{aligned}
\end{equation} we find that the classical bound is saturated by the distribution \begin{equation}
    P(a_1a_2a_3|x_1x_2x_3) = \delta(a_1,x_1)\delta(a_2,x_2)\delta(a_3,x_3).
\end{equation}
This distribution can be easily realized by observables $\sigma_0=I,\sigma_1=-I$ and any TI product state. If we choose \begin{equation}\label{ex1}
    \sigma_0=\begin{pmatrix}
        1 & 0\\
        0 & -1
    \end{pmatrix}, \sigma_1=\begin{pmatrix}
        -1 & 0\\
        0 & 1
    \end{pmatrix},
\end{equation}
the state should be restricted to $\ket{000}$.

In general, to construct a 322-type Hamiltonian that achieves the classical bound $\mathcal{L}$, one should first find a TI classical distribution $P(a_1a_2a_3|x_1x_2x_3)$ that can saturate the classical bound, and then design a Hamiltonian to produce $P$. We explain how to find $P$ and analyze the properties of $P$. 

Any convex combination of $S_Q$ can lead to the classical bound, and we first consider a vertex $Q$. If $Q$ is both TI and LD, we simply set the operators to $\{I,-I\}$ depending on the deterministic pattern of $Q$, and the state is irrelevant. For example, considering the first 322-type model in Table~\ref{tab:322}, the simplest $Q$ is $Q(010101)=1$, and we have $\sigma_0=I,\sigma_1=-I$. If one uses other observables instead of $\{I,-I\}$, we should properly choose the observables and states such that the statistics $P$ can be recovered, as we did in Eq.~\eqref{ex1}.

The interesting part is the case when $Q$ is a convex combination of non-TI LD distributions. Let $Q=\sum_{i=1}^{r}\lambda_i Q_i$, where each $Q_i$ is LD and corresponds to $P_i$ by Eq.~\eqref{finetheorem}. The strategy is to find two observables $\sigma_0,\sigma_1$ and $r$ states such that the $i$-th state with $\sigma_0,\sigma_1$ can produce $P_i$. Consider the inequality Eq.~\eqref{322-model1}, and we choose the vertex $Q=\sum_{i=1}^{4}\lambda_i Q_i$ as the LHV model, where \begin{equation}
\begin{aligned}
        &Q_1(010110) = 1,
        Q_2(011010) = 1,\\
        &Q_3(101001) = 1,
        Q_4(100101) = 1.
\end{aligned}
\end{equation}
For each $Q_i$, by assigning $01\rightarrow \ket{0}, 10\rightarrow \ket{1}$, we choose the states in the form of\begin{equation}
\begin{aligned}
        & \ket{\phi_1} = \ket{001},
        \ket{\phi_2} = \ket{011},
        \\ & \ket{\phi_3} = \ket{110},
        \ket{\phi_4} = \ket{100} .   
\end{aligned}
\end{equation}
To recover $P_i$ with $\ket{\phi_i}$, the observables should be \begin{equation}
    \sigma_0=\begin{pmatrix}
        1 & 0\\
        0 & -1
    \end{pmatrix}, \sigma_1=\begin{pmatrix}
        -1 & 0\\
        0 & 1
    \end{pmatrix}.
\end{equation}
To see why the observables fit, one can check  whether each state $\ket{\phi_i}$ can produce $P_i$. For example, $P_1$ has the property that if the first party is measured with observable $\sigma_0$ ($\sigma_1$), the outcome must be $0$ ($1$), which is consistent with measuring $\ket{\phi_1}$ using $\sigma_0$ and $\sigma_1$. The other two parties follow the same reasoning.

Since the set of $Q_i$ admits an irreducible domino loop $\{010110,011010,101001,100101\}$, we know that the states $\ket{\phi_i}$ also form a periodic state, i.e., \begin{equation}
    \ket{\Phi}=\ket{0011}^{\otimes \infty}.
\end{equation}
This is a state of period 4, which is exactly the size of the domino loop. The symmetrization of $\ket{\Phi}$ gives the one-site TI state. The above construction method is general. Indeed, for any vertex $Q=\sum_i\lambda_iQ_i$, we know that the set of deterministic components $Q_i$ always forms a domino loop. This loop gives the structure of the ground states, and one only needs to find suitable observables to produce $Q_i$. The periodic structure of the ground states is the same as that of the domino loop.

Some different vertices in $S_Q$ may correspond to a single Hamiltonian $H$, which leads to a degenerate ground space consisting of multiple distinct periods. For instance, let $\sigma_0,\sigma_1$ be \begin{equation}\label{322-obs-multiperiods}
	\sigma_0 = \begin{pmatrix}
		1 &  0 & 0\\ 
		0 & -1 & 0\\
        0 &  0 & 1
	\end{pmatrix}, \quad
	\sigma_1 = 
	\begin{pmatrix}
		1 & 0  &0 \\ 
	   0 & -1  &0 \\
               0 &         0  &-1
	\end{pmatrix}.
\end{equation}
We list the vertices with distinct periods and corresponding ground states as below. \begin{enumerate}\label{numbered-list}
    \item For $Q(010101)=1$, the ground state is $\ket{222}^{\otimes \infty}$.
    \item For $Q(110101)=Q(010111)=Q(011101)=1/3$, the ground state is $\ket{122}^{\otimes \infty}$.
    \item For $Q(001101)=Q(110101)=Q(010100)=Q(010011)=1/4$, the ground state is $\ket{0122}^{\otimes \infty}$.
    \item For $Q(001101)=Q(110101)=Q(010111)=Q(011100)=Q(110011)=1/5$, the ground state is $\ket{01221}^{\otimes \infty}$.
    \item For $Q(000011)=Q(001101)=Q(110101)=Q(010111)=Q(011100)=Q(110000)=1/6$, the ground state is $\ket{001221}^{\otimes \infty}$.
\end{enumerate}

In fact, the TI ground sector of $H$ can be characterized using the domino-loop construction. By the TI ground sector, we mean the set of energy-density-minimizing states within the class of TI or $p$-periodic states; symmetrizing a 
$p$-periodic state produces a TI state. Although the full ground-state manifold may in general include non-TI states, we restrict attention to this TI/
$p$-periodic sector. Since $H$
is diagonal in the computational basis, all basis product states
$\{\ket{\cdots s_1 s_2 s_3 \cdots}\}$ are exact eigenstates of $H$. The problem therefore reduces to
identifying those periodic eigenstates that attain the minimal eigenvalue. We begin with the vertex set $S_Q$, namely all LTI-$Q$ vertices that attain the classical bound via
Eqs.~\eqref{322-expectation} and~\eqref{finetheorem}. For each vertex in $S_Q$, the associated irreducible domino loop directly specifies a periodic computational-basis state; these
states are manifestly ground states of $H$. However, this does not yet exhaust the TI ground sector:
there also exist ground states whose underlying domino loops are \emph{reducible}, i.e., built from
convex combinations of distinct vertices in $S_Q$.

A concrete example is the $7$-period state
$\ket{\Psi}=\ket{0122122}^{\otimes \infty}$. It is a ground state of $H$, but the corresponding LHV
model $Q$ is a convex combination of the second and third vertices listed above, and hence is not
irreducible. Importantly, $\ket{0122122}^{\otimes \infty}$ cannot be generated as a linear
combination of $\ket{122}^{\otimes \infty}$ and $\ket{0122}^{\otimes \infty}$, so it lies outside
the span of ground states obtained solely from irreducible domino loops. To capture the ground states systematically, let
$\ket{\Phi}=\ket{s_1 s_2 \cdots s_r}^{\otimes \infty}$ be an arbitrary $r$-periodic computational-basis
state, and let $Q_{1,2,\ldots,r}^{\otimes \infty}$ denote the unique deterministic LHV model reproducing the
statistics generated by $\{\sigma_0,\sigma_1\}$ on $\ket{\Phi}$. Consider the TI symmetrization of
its three-site marginals,
\begin{equation}
\overline{Q}_{1,2,3}
=\frac{1}{r}\sum_{i=1}^{r} Q_{i,i+1,i+2},
\end{equation}
where indices are taken modulo $r$ and $Q_{i,i+1,i+2}$ is the marginal of $Q_{1,2,\ldots,r}$ on
sites $(i,i+1,i+2)$. It is clear that $\ket{\Phi}$ is a ground state of $H$ if and only if $\overline{Q}_{1,2,3}$ reproduces the classical bound through Eq.~\eqref{finetheorem} and Eq.~\eqref{322-expectation}. The key point is that $\overline{Q}_{1,2,3}$ reproduces the classical bound if and only if
$\overline{Q}_{1,2,3}$ lies in the convex hull of $S_Q$ (since the set of LHV models is a convex polytope, all LHV models attaining the classical bound must lie on the corresponding facet, which is determined by $S_Q$). This criterion includes the irreducible-loop ground states
as a special case and also accounts for reducible constructions such as
$\ket{0122122}^{\otimes\infty}$.

\section{Semidefinite-program framework for TI systems and finite-size reductions}
\label{sec:npa}
In this section, we demonstrate how to use the NPA hierarchy to derive tight lower bounds on the 322-type TI contextuality witnesses (the same construction applies naturally to the 232-type case), and to derive the state structure of the optimal ground states. Since we are dealing with TI systems, additional treatments are required. The difficulty is that the allowed states must be TI reduced states, but an exact characterization of TI reduced states using semidefinite programming (SDP) is not available in closed form \cite{blakajSetReducedStates2024}. Therefore, we consider an outer relaxation based on LTI states, noting that LTI is a necessary condition for a state to be TI. This relaxation becomes progressively tighter as the subsystem size increases, and it coincides with the TI set in the infinite-size limit. 

Furthermore, we find that the NPA hierarchy with periodic boundary conditions (PBC) reproduces the quantum limit when the system size equals $p_{\mathcal{Q}}$. This observation naturally leads to finite PBC inequalities, and we prove that these PBC inequalities share the same quantum and classical bounds as their TI counterparts. Exploiting this PBC equivalence, we show how to derive exact solutions for the 322-type inequalities and present two explicit examples. Finally, we highlight the structural properties of the optimal ground states.

\subsection{The NPA hierarchy with LTI}\label{subsec:lti-npa}
In general, obtaining the quantum limit of an inequality is highly non-trivial, as one must prove that the achieved value is optimal over all possible quantum states and measurement settings. In practice, this requires deriving a tight lower bound and identifying a state and observables that saturate it. To determine such lower bounds, one commonly employs the NPA hierarchy \cite{navascuesBoundingSetQuantum2007,navascuesConvergentHierarchySemidefinite2008}. 

Assuming an $m$-partite inequality of the form $E = \langle\mathcal{E}\rangle \geq \mathcal{L}$, we describe how to derive lower bounds on $E$ using the NPA hierarchy. Consider a Bell scenario with $m$ parties, each having $X$ measurement inputs, and each input producing two possible outcomes ($\pm1$). Let $\sigma_{a_i}^{(i)}$ denote the operator acting on site $i$, where $a_i \in \{0, 1, \dots, X-1\}$. These operators satisfy $(\sigma_{a_i}^{(i)})^2 = 1$ (the identity operator) and commute across different sites. The objective function $E$, specified by the inequality, is the expectation value of a noncommutative  polynomial $\mathcal{E}$. 

Let $(\mathcal{Q}_i)_i$ ($i = 1, 2, \dots, t$) be a generating sequence of monomials of the form
\[
\mathcal{Q}_i = \sigma_{a_1}^{(i_1)} \sigma_{a_2}^{(i_2)} \cdots \sigma_{a_k}^{(i_k)},
\]
where the level of the NPA hierarchy refers to the maximum degree among these monomials. The sequence $(\mathcal{Q}_i)_i$ should satisfy the condition that $\mathcal{E}$ can be written as a linear combination of $\mathcal{Q}_i^{\dagger}\mathcal{Q}_j$. 

We define the matrix $\mathcal{O}$ with components $\mathcal{O}_{i,j} = \mathcal{Q}_i^\dagger \mathcal{Q}_j$. Let $(M_k)_k$ ($k = 1, 2, \dots, s$) be a basis for the vector space spanned by components of $\mathcal{O}$. The basis is chosen such that $M_1 = 1$ and each term is Hermitian. The objective operator $\mathcal{E}$ can then be expressed as
\[
\mathcal{E} = \sum_{k=1}^s c_k M_k,
\]
and $\mathcal{O}$ can be written as
\[
\mathcal{O} = \sum_{k=1}^s A_k M_k,
\]
where $A_k$ are $t \times t$ Hermitian matrices. Let $x_k = \langle M_k \rangle$ denote the expectation value of $M_k$. The moment matrix and objective function are then
\[
\Gamma = \sum_{k=1}^s A_k x_k, \quad E = \sum_{k=1}^s c_k x_k.
\]

By construction, the moment matrix $\Gamma$ is positive semidefinite, and the quantum limit of the inequality can be relaxed to a SDP in the form of
\begin{equation} \label{npa-org}
    \begin{aligned} 
    \text{min} \quad & \sum_{k=1}^s c_k x_k \\
    \text{subject to} \quad & x_1 = 1, \quad
\Gamma = \sum_{k=1}^s A_k x_k \succeq 0.
\end{aligned}
\end{equation}
Eq.~\eqref{npa-org} always produces a lower bound on the quantum limit, which becomes progressively tighter as more monomials and higher levels of the hierarchy are included. Although it is generally unknown whether a specific level is sufficient, in practice, levels up to five are typically enough to reproduce the quantum limit. However, since only TI reduced states are physically admissible for the considered TI inequalities, a careful treatment of translation invariance is necessary. 

Assume that the $m$-partite TI inequality is $\langle\mathcal{E}\rangle_{\text{TI}}\geq \mathcal{L}$. To perform the NPA hierarchy with LTI, we use a system of $n$ parties with $n\geq m$, while the objective function is acting on the first $m$ sites. To enforce LTI, we simply impose additional constraints in the form of 
\begin{equation}\label{eq:lti}
    \langle \sigma_{a_1}^{(i_1)}\sigma_{a_2}^{(i_2)}\cdots \sigma_{a_k}^{(i_k)}\rangle = \langle \sigma_{a_1}^{(i_1+1)}\sigma_{a_2}^{(i_2+1)}\cdots \sigma_{a_k}^{(i_k+1)}\rangle,
\end{equation}
where $i_1\leq i_2 \leq \cdots \leq i_k \leq n-1$.
We denote these constraints by $Lx=0$. Then, the LTI-NPA is given by the SDP:
\begin{equation}
	\label{lti-npa}
	\begin{aligned}
		\text{min} \quad&\sum_{l=1}^s c_{l} x_l \\
		\text{s.t.}\quad &   x_1=1,  \textstyle\sum_{l=1}^s A_{l} x_l \succeq 0, L x = 0.\\
	\end{aligned}
\end{equation}
The optimal value of Eq.~\eqref{lti-npa} always provides a lower bound on the quantum limit of $\langle \mathcal{E} \rangle_{\TI}$. This bound becomes progressively tighter as the number of parties $n$ is increased and higher levels of the hierarchy are employed. 
The quantum limits $\mathcal{Q}$ reported in Table~\ref{tab:322} can be reproduced from Eq.~\eqref{lti-npa} using suitably chosen values of $n$ and hierarchy levels.

\subsection{The finite PBC inequalities induced by TI inequalities} \label{subsec:pbc-reduction}
Since the optimal states are periodic, it is natural to employ the NPA hierarchy with periodic boundary conditions (PBC), even though a TI state on a finite ring does not, in general, correspond to a TI state in the thermodynamic limit. There are two possible ways to implement the PBC version of the NPA hierarchy, referred to as PBC-NPA. One is to impose periodic constraints on the moment matrix $\Gamma$ in the form of \begin{equation}
    \langle\sigma_{a_1}^{(i_1)}\sigma_{a_2}^{(i_2)}\cdots \sigma_{a_k}^{(i_k)}\rangle = \langle \sigma_{a_1}^{(i_1+1)}\sigma_{a_2}^{(i_2+1)}\cdots \sigma_{a_k}^{(i_k+1)}\rangle,
\end{equation}
where $i_1\leq i_2 \leq \cdots \leq i_k \leq n$, and $i_k+1 =1$ if $i_k=n$. The other way is to simply switch the objective function from the local term $\langle\mathcal{E}\rangle$ to the average of the translated local terms. For example, consider a 322-type Bell functional in the form of $
\sum_{x=0}^{1}J_{x}\langle\sigma^{(1)}_x\rangle+\sum_{x,y=0}^{1}J_{xy}\langle\sigma^{(1)}_x\sigma^{(2)}_y\rangle
+\sum_{x,z=0}^{1}J_{xz}\langle\sigma^{(1)}_x\sigma^{(3)}_z\rangle,$ the translated average is \begin{equation}\label{pbc-obj}
\begin{aligned}
    \frac{1}{n}\sum_{i=1}^{n}(&\sum_{x}J_{x}\langle\sigma^{(i)}_x\rangle+\sum_{x,y}J_{xy}\langle\sigma^{(i)}_x\sigma^{(i+1)}_y\rangle
\\
+&\sum_{x,z}J_{xz}\langle\sigma^{(i)}_x\sigma^{(i+2)}_z\rangle).
\end{aligned}
\end{equation}
The two ways are equivalent, because one can construct solutions from each other. Remarkably, for all ten inequalities in Table~\ref{tab:322}, the PBC-NPA with $p_{\mathcal{Q}}$ sites reproduces the quantum limits of the TI contextuality witnesses. This implies that the optimal TI states defined on a finite ring may admit genuine TI extensions to the thermodynamic limit. When the PBC-NPA already attains the quantum limit, finite PBC inequalities arise naturally from their TI counterparts.

Consider the 322-type TI inequality in the form of
\begin{equation}\label{322-org}
\begin{aligned}
\langle\mathcal{E}\rangle_{\text{TI}}=\Bigl\langle &\sum_{x=0}^{1}J_x\sigma^{(1)}_x
+\sum_{x,y=0}^1J_{xy}\sigma^{(1)}_x\sigma^{(2)}_y \\+&\sum_{x,z=0}^1J_{xz}\sigma^{(1)}_x\sigma^{(3)}_z\Bigl\rangle_{\text{TI}}
   \geq \mathcal{L},
\end{aligned}
\end{equation} and we assume that the PBC-NPA with $p$ sites gives the quantum limit $\mathcal{Q}$. We show that the induced finite PBC inequality
\begin{equation}\label{322-pbc}
\begin{aligned}
    \langle\mathcal{E}_{\PBC}\rangle=\frac{1}{p}\sum_{i=1}^{p}(&\sum_{x=0}^{1}J_{x}\langle\sigma^{(i)}_x\rangle+\sum_{x,y=0}^{1}J_{xy}\langle\sigma^{(i)}_x\sigma^{(i+1)}_y\rangle\\+&\sum_{x,z=0}^{1}J_{xz}\langle\sigma^{(i)}_x\sigma^{(i+2)}_z\rangle) \geq \mathcal{L}
\end{aligned}
\end{equation}
is a valid contextuality witness, and the quantum limit is also $\mathcal{Q}$.

The optimal value given by the PBC-NPA of the TI inequality in Eq.~\eqref{322-org} coincides with that of the NPA of the finite inequality in Eq.~\eqref{322-pbc}. Consequently, $Q$ serves as a lower bound of $\langle \mathcal{E}_{\PBC}\rangle$. To saturate this bound, we use the same optimal observables as in Eq.~\eqref{322-org} and the corresponding ground states. It then remains to show that $\mathcal{L}$ represents the classical bound of $\langle\mathcal{E}_{\PBC}\rangle$.

By Fine's theorem, any classical probability distribution $P(a_1a_2a_3|x_1x_2x_3)$ admits a LHV model $Q(a_{1,0},a_{1,1},a_{2,0},a_{2,1},a_{3,0},a_{3,1})$. Treating $P,Q$ as two vectors, there exists a projection $M$ such that $P=MQ$. $\langle\mathcal{E}_{\PBC}\rangle$ ($\langle\mathcal{E}\rangle_{\text{TI}}$) is a linear function on $P$, i.e., there exists a vector $f$ ($h$) such that $\langle\mathcal{E}_{\PBC}\rangle=f^TP$ ($\langle\mathcal{E}\rangle_{\text{TI}}=h^TP$). Then, the classical bound of $\langle\mathcal{E}_{\PBC}\rangle$ is equal to the optimal value of the linear programming in the form of \begin{equation}
	\label{322-cb}
	\begin{aligned}
		\text{min} \quad&f^TP_p \\
		\text{s.t.}\quad &   P_p=M_pQ_p, \quad \sum_i Q_{p,i}=1,\quad Q_{p,i}\geq 0\\
	\end{aligned}
\end{equation}
where $P_p$ and $Q_p$ denote the distributions on $p$ parties, and $M_p$ is the projection.
On the other hand, the classical bound of $\langle\mathcal{E}\rangle_{\text{TI}}$ is given by  \begin{equation}
	\label{322-ti-cb}
	\begin{aligned}
		\text{min} \quad&h^TP_3 \\
		\text{s.t.}\quad &   P_3=M_3Q_3, \quad \sum_i Q_{3,i}=1, \\
        & Q_{3,i}\geq 0, \quad Q_3 \text{ is TI.}
	\end{aligned}
\end{equation}
Demanding that $Q_3$ is TI is equivalent to enforcing LTI on $Q_3$, i.e., \begin{equation}\label{lti-q}
    \begin{aligned}
        &\sum_{a_{3,0},a_{3,1}}Q_3(a_{1,0},a_{1,1},a_{2,0},a_{2,1},a_{3,0},a_{3,1}) \\
        = &\sum_{a_{1,0},a_{1,1}}Q_3(a_{1,0},a_{1,1},a_{2,0},a_{2,1},a_{3,0},a_{3,1}).
    \end{aligned}
\end{equation}

As in the two formulations of PBC-NPA, we can switch the objective function in Eq.~\eqref{322-cb} to the local term $\langle\mathcal{E}\rangle$ and for $i=2,\ldots,p$, demand $P$ to satisfy \begin{equation}
\begin{aligned}
    &P_p(a_1a_2\cdots a_p|x_1x_2\cdots x_p)\\=&P_p(a_ia_{i+1}\cdots a_{i+p-1}|x_ix_{i+1}\cdots x_{i+p-1}),
\end{aligned}
\end{equation}
with all site indices understood modulo $p$.
This is equivalent to demanding that for $i=2,\ldots,p$ \begin{equation}\label{pbc-q}
\begin{aligned}
    &Q_p(a_{1,0},a_{1,1},a_{2,0},a_{2,1},\ldots, a_{p,0},a_{p,1})\\=&Q_p(a_{i,0},a_{i,1},a_{i+1,0},a_{i+1,1},\ldots,a_{i+p-1,0},a_{i+p-1,1}).
\end{aligned}
\end{equation}
Eq.~\eqref{322-cb} is equivalent to \begin{equation}
	\label{322-cb1}
	\begin{aligned}
		\text{min} \quad&h^TP_3 \\
		\text{s.t.}\quad &   P_3=\text{Tr}_{4,\ldots,p}(P_p),\quad P_p=M_pQ_p, \\ & \sum_i Q_{p,i}=1,\quad Q_{p,i}\geq 0,\\& Q_p \text{ satisfies Eq.~\eqref{pbc-q}}.
	\end{aligned}
\end{equation}

Let $\mathcal{L}_{\TI}$ denote the optimal value of Eq.~\eqref{322-ti-cb}, and let $\mathcal{L}_{p}$ denote the optimal value of Eq.~\eqref{322-cb1}. Since Eq.~\eqref{lti-q} is a necessary condition for Eq.~\eqref{pbc-q}, we have $\mathcal{L}_{p}\geq \mathcal{L}_{\TI}$.
Moreover, if Eq.~\eqref{322-ti-cb} admits an optimal vertex solution associated with a domino loop of size $p$, then in fact
\[
\mathcal{L}_{p}=\mathcal{L}_{\TI}.
\]
The reason is that such a vertex solution can be lifted to a feasible solution of Eq.~\eqref{322-cb1} with the same objective value. Concretely, let $Q_3^*=\sum_{i=1}^{p}\lambda_i Q_i$ be an optimal vertex solution of Eq.~\eqref{322-ti-cb} admitting a domino loop of size $p$, where
\begin{equation} \label{eq:optimal-sol-size-p}
    Q_3^*(y_{i,1},y_{i,2},y_{i,3},y_{i,4},y_{i,5},y_{i,6})=1/p,i=1,\ldots,p.
\end{equation}
The corresponding domino loop is specified by the bit sequence
\[
\{\vec{y}_i=(y_{i,1},y_{i,2},y_{i,3},y_{i,4},y_{i,5},y_{i,6})\}_{i=1}^{p},
\]
satisfying $(y_{i,3},y_{i,4},y_{i,5},y_{i,6})
=
(y_{i+1,1},y_{i+1,2},y_{i+1,3},y_{i+1,4})$ for $i=1,\ldots,p,$
with indices understood modulo $p$. We then construct a new bit string $\vec{z}$ by concatenating $\vec{y}_1$ with the last two bits of each $\vec{y}_i$ for $i=2,\ldots,p-2$, namely \begin{equation}\label{eq:zfromy}
\begin{aligned}
\vec{z}=(y_{1,1},y_{1,2},y_{1,3},y_{1,4},y_{1,5},y_{1,6},y_{2,5},y_{2,6},\ldots, \\y_{p-2,5},y_{p-2,6}).
\end{aligned}
\end{equation}
Let $C$ denote the operator that cyclically shifts $\vec{z}$ by two bits. We define $Q_p^*$ by
\begin{equation}\label{eq:symmetrizing-z}
    Q_p^*(C^j(\vec{z}))=\frac{1}{p},
\qquad j=0,\ldots,p-1.
\end{equation}
By construction, $Q_p^*$ satisfies Eq.~\eqref{pbc-q}. Moreover, it is straightforward to verify that $Q_p^*$ yields the same objective value as $Q_3^*$. Therefore, $\mathcal{L}_{p}\leq \mathcal{L}_{\TI}$, and hence $\mathcal{L}_{p}=\mathcal{L}_{\TI}$.

For the witnesses listed in Table~\ref{tab:322}, the required classical finite-ring
solutions can be constructed explicitly from domino loops at the
corresponding quantum-selected periods. The resulting loops are shown
in Table~\ref{tab:domino-loops}. This provides a direct check that the finite PBC inequalities
used here have the same classical bound as their infinite TI counterparts.
The classical bound of the 232-type inequality is obtained in the same
way, with the period-three domino loop $(100111,111101,101100)$.
\begin{table*}[htbp]
    \caption{The domino loop of size $p_{\mathcal{Q}}$ corresponding to an optimal vertex solution for each model in Table~\ref{tab:322}. The sequence $\{\vec{y}_i\}$ specifies the domino loop, and the associated vertex solution is obtained from Eq.~\eqref{eq:optimal-sol-size-p}. The bit string $\vec{z}$ is constructed according to Eq.~\eqref{eq:zfromy}, and the corresponding optimal solution $Q_p^*$ is obtained from Eq.~\eqref{eq:symmetrizing-z}.}
    \centering
    \renewcommand\arraystretch{1.5}
    \setlength{\tabcolsep}{2.2mm}{
    \begin{tabular}{|c|c|c|}	
    \hline
    No. &  the domino loops $\{\vec{y}_i\}$ & $\vec{z}$
    \\ \hline						
    1&	010110,011010,101001,100101 & 01011010
    \\ \hline									 	 
    2&	000101,010110,011000,100001 & 00010110
    \\	\hline								 	 	
    3&	010110,011010,101001,100101 & 01011010
    \\ \hline
    4&	000001,000101,010100,010000 & 00000101
    \\ \hline				 	
    5& 	000101,010110,011000,100001 & 00010110
    \\ \hline									 		
    6&	000100,010011,001110,111000,100001 & 0001001110
    \\ \hline
    7&	000111,011100,110010,001000,100001 & 0001110010
    \\ \hline
    8&	000100,010010,001011,101100,110001 & 0001001011
    \\ \hline
    9&	000001,000110,011000,100000	& 00000110
    \\ \hline	
    10&	000001,000111,011101,110100,010000 & 0000011101
\\ \hline
    \end{tabular}}
    \label{tab:domino-loops}
\end{table*}

Having obtained a finite PBC inequality induced by a TI witness, we next address its \emph{tightness}. An inequality is tight if the set of classical points saturating it 
forms a facet of the classical convex polytope, i.e., a face of dimension one less than that of the polytope. If the saturating set has a lower dimension, the inequality defines a lower-dimensional face. Tightness is operationally useful: a facet-defining inequality is nonredundant and probes a maximal-dimensional face of the relevant classical polytope. Since tightness depends on the translation-invariance constraints imposed on the finite ring, we consider two settings. In the first case, we impose no symmetry on the finite ring. In this case, the induced PBC inequalities are not tight: the faces defined by the inequalities do not attain the dimension of a facet. Equivalently, without imposing TI, the induced PBC inequalities define faces rather than facets.

In the second setting, we impose TI on the ring. Under this TI constraint, the induced PBC inequalities define faces of the TI classical polytope; moreover, in several instances they are tight. Examples include the 232-type inequality Eq.~\eqref{232-pbc} and the fourth inequality in Table~\ref{tab:322}, namely,\begin{equation}
    \begin{aligned}
        \frac{1}{4}\sum_{i=1}^{4}\langle&-2\sigma_0^{(i)} -2 \sigma_1^{(i)} -2 \sigma_0^{(i)}\sigma_0^{(i+1)}+\sigma_0^{(i)}\sigma_1^{(i+1)}\\&-\sigma_1^{(i)}\sigma_0^{(i+1)}-2\sigma_1^{(i)}\sigma_1^{(i+1)}+\sigma_0^{(i)}\sigma_0^{(i+2)}\\&+2\sigma_1^{(i)}\sigma_0^{(i+2)}+\sigma_1^{(i)}\sigma_1^{(i+2)}\rangle \geq -4.
\end{aligned}
\end{equation}

\subsection{Exact solutions derived from the PBC equivalence}\label{subsubsec-322-exact}
We show how to derive analytical solutions for the 322-type inequalities by utilizing the observation that TI inequalities are equivalent to PBC inequalities. Consider the third inequality in Table~\ref{tab:322} in the form of \begin{equation}
    \begin{aligned}
        &\langle-3\sigma_0^{(1)} + \sigma_1^{(1)} + \sigma_0^{(1)}\sigma_0^{(2)}+\sigma_0^{(1)}\sigma_1^{(2)}+\sigma_1^{(1)}\sigma_0^{(2)}\\&-\sigma_1^{(1)}\sigma_1^{(2)}+\sigma_0^{(1)}\sigma_0^{(3)}-\sigma_1^{(1)}\sigma_0^{(3)}+\sigma_1^{(1)}\sigma_1^{(3)}\rangle \geq -3,
\end{aligned}
\end{equation} and the parameterization is the same as Eq.~\eqref{322 parameterized observables}. That is, $\sigma_0,\sigma_1$ are parameterized as \begin{equation}\footnotesize
	\sigma_0 = \begin{pmatrix}
		1 &  0 & 0\\ 
		0 & -1 & 0\\
        0 &  0 & 1
	\end{pmatrix},
	\sigma_1 = 
	\begin{pmatrix}
		\cos(2\theta) & -\sin(2\theta)  &0 \\ 
	   -\sin(2\theta) & -\cos(2\theta)  &0 \\
               0 &         0  &-1
	\end{pmatrix},
    \label{322 parameterized observables model3}
\end{equation}
and the Hamiltonian $H(\theta)$ is given by \begin{equation}\label{322-model3}
    \begin{aligned}
    H=\sum_{i=1}^{\infty} &-3\sigma_0^{(i)} + \sigma_1^{(i)} + \sigma_0^{(i)}\sigma_0^{(i+1)}+\sigma_0^{(i)}\sigma_1^{(i+1)}\\&+\sigma_1^{(i)}\sigma_0^{(i+1)}-\sigma_1^{(i)}\sigma_1^{(i+1)}+\sigma_0^{(i)}\sigma_0^{(i+2)}\\&-\sigma_1^{(i)}\sigma_0^{(i+2)}+\sigma_1^{(i)}\sigma_1^{(i+2)}.
\end{aligned}
\end{equation}

The optimal numerical solution for $\theta_{opt}$ is $0.78524585$, and the goal is to derive an exact solution for $\theta_{opt}$ and the ground states.
By the PBC equivalence, the ground-state energy density of Eq.~\eqref{322-model3} (using $\theta_{opt}$) is equivalent to that of the PBC Hamiltonian \begin{equation}\label{322-model3-pbc}
    \begin{aligned}
    H_{\PBC}=\sum_{i=1}^{4} &-3\sigma_0^{(i)} + \sigma_1^{(i)} +\sigma_0^{(i)}\sigma_0^{(i+1)}+\sigma_0^{(i)}\sigma_1^{(i+1)}\\&+\sigma_1^{(i)}\sigma_0^{(i+1)}-\sigma_1^{(i)}\sigma_1^{(i+1)}+\sigma_0^{(i)}\sigma_0^{(i+2)}\\&-\sigma_1^{(i)}\sigma_0^{(i+2)}+\sigma_1^{(i)}\sigma_1^{(i+2)}.
\end{aligned}
\end{equation}

The problem now becomes how to find $\theta_{opt}$ such that the ground-state energy density of $H_{\PBC}$ is the lowest among all possible $\theta$. $H_{\PBC}$ has a dimension of $81\times 81$, which is not convenient for analytically computing the eigenvalues, so we use translation symmetry to further block diagonalize $H_{\PBC}$. Since $H_{\PBC}$ is invariant under translations on a ring of four sites, the one-site translation operator $T$ commutes with $H_{\PBC}$, i.e., $[H_{\PBC},T]=0$. The eigenvalues of $T$ are $\{1,-1,i,-i\}$, which correspond to four invariant eigenspaces. Let $\Pi_1,\Pi_2,\Pi_3,\Pi_4$ be the four projectors onto the four eigenspaces. Since $[H_{\PBC},T]=0$, we can use $\Pi_1,\Pi_2,\Pi_3,\Pi_4$ to block diagonalize $H_{\PBC}$ into four blocks, which have sizes $21,18,18,24$. We compute the exact eigenvalues of each block, which are analytical functions of $\theta$, and minimize the smallest eigenvalue with respect to $\theta$. 

As it turns out, the optimal $\theta_{opt}$ is $\frac{\pi}{4}$, and the lowest energy density is $-\frac{5+\sqrt{2}}{2}$. Substituting $\theta_{opt}$ into Eq.~\eqref{322 parameterized observables model3}, we have 
\begin{equation}
	\sigma_0 = \begin{pmatrix}
		1 &  0 & 0\\ 
		0 & -1 & 0\\
        0 &  0 & 1
	\end{pmatrix}, \quad
	\sigma_1 = 
	\begin{pmatrix}
		0 & -1  &0 \\ 
	   -1 & 0  &0 \\
     0 &  0  &-1
	\end{pmatrix},
\end{equation}
and the ground state is \begin{equation}
    \ket{\phi}=(\alpha\ket{\Phi^-}+\beta\ket{\Psi^+})\otimes \ket{22},
\end{equation}
where $\alpha=\frac{\sqrt{2-\sqrt{2}}}{2},\beta=\frac{\sqrt{2+\sqrt{2}}}{2}$ and $\ket{\Phi^-}=\frac{1}{\sqrt{2}}(\ket{00}-\ket{11}), \ket{\Psi^+}=\frac{1}{\sqrt{2}}(\ket{01}+\ket{10})$. 

The first two sites of $\ket{\phi}$ are entangled, but they do not entangle with the last two sites. This particular structure ensures that the ground state of $H_{\PBC}$ corresponds to an infinite TI state. In fact, $\ket{\phi}^{\otimes \infty}$ is the exact ground state of the infinite Hamiltonian $H$ in Eq.~\eqref{322-model3}, and $-\frac{5+\sqrt{2}}{2}$ is the exact ground-state energy density of $H$. The ground space of $H$ is degenerate, and $\ket{\phi}^{\otimes \infty}$ with the other three translated states of $\ket{\phi}^{\otimes \infty}$ span the ground space. If one uses uMPS to compute the ground state, the symmetrization of $\ket{\phi}^{\otimes \infty}$ will be returned, and the transfer matrix has eigenvalues $\{1,-1,i,-i\}$. 

For the other inequalities in Table~\ref{tab:322}, we can use the same method to derive exact solutions. Most of the solutions are cumbersome, involving roots of high-order polynomial equations, so we do not show the detailed results here. For the second inequality in Table~\ref{tab:322} in the form of 
\begin{equation}
\begin{aligned}
\bigl\langle\,
&-4\sigma_0^{(1)} + 2\sigma_1^{(1)}
+ 2\sigma_0^{(1)}\sigma_0^{(2)}
+ 2\sigma_0^{(1)}\sigma_1^{(2)} \\
&+ 2\sigma_1^{(1)}\sigma_0^{(2)}
- 4\sigma_1^{(1)}\sigma_1^{(2)}
+ \sigma_0^{(1)}\sigma_0^{(3)}
- \sigma_0^{(1)}\sigma_1^{(3)} \\
&- \sigma_1^{(1)}\sigma_0^{(3)}
+ 3\sigma_1^{(1)}\sigma_1^{(3)}
\,\bigr\rangle \ge -6 .
\end{aligned}
\end{equation}
Using the same parameterization as in Eq.~\eqref{322 parameterized observables model3}, we find that the optimal $\theta$ is $\arctan(\sqrt{\frac{5}{3}})$, and the exact quantum limit is $-\frac{3}{4}(6+\sqrt{6})$.

\subsection{The structure of the ground states that achieve maximal quantum violations}
For the studied TI contextuality witness families, the ground state $\ket{\Psi}$ (as a uMPS) of $H$ that achieves the quantum limit has a simple structure that is related to the period $p$ of $\ket{\Psi}$. That is, $\ket{\Psi}$ is the symmetrization of a $p$-periodic state. When the violation is maximal, the state has an additional $p$-block product state structure.

Let $\ket{\Psi(A)}$ be the ground state that saturates the quantum limit. As the transfer matrix $E_{A}$ shows a period $p_{\mathcal{Q}}>1$, the state admits a periodic decomposition \cite{MPS-representation}. Let $U$ be the unitary eigenoperator corresponding to the eigenvalue $\omega=e^{2\pi i/p_{\mathcal{Q}}}$. The spectrum decomposition of $U$ is $U=\sum_k\omega^kP_k$, where $P_kP_j=\delta_{k,j}P_k$ and $\sum_k P_k=\mathbb{I}$. Define the site-dependent tensors $A^{s_j}=P_{j}A^sP_{j+1}$ for $j=1,\ldots,p_{\mathcal{Q}}$ and the unit cell $\mathbf{A^{\mathbf{s}}}=A^{s_1}\cdots A^{s_{p_\mathcal{Q}}}$. Let $\ket{\Psi_0}$ be the state generated by $\mathbf{A^{\mathbf{s}}}$. Then, $\ket{\Psi}$ can be written as \begin{equation}
    \ket{\Psi}=\frac{1}{\sqrt{p_{\mathcal{Q}}}}\sum_{k=0}^{p_{\mathcal{Q}}-1}T^{k}(\ket{\Psi_0}),
\end{equation}
where $T$ is the translation operator by one site.

When the ground state $\ket{\Psi(A)}$ reaches the maximal quantum violation, we find that the nonzero spectrum of $E_A$ lies exactly on the unit circle (all the nonzero eigenvalues have magnitude 1). Also, for the studied
232/322 TI witness families, we observe that there is always a rank-1 projector $P_i$ that makes the state block-product. Specifically, the optimal state takes the form of $\ket{\Psi_{gs}}=\ket{\psi_{p_{\mathcal{Q}}}}^{\otimes \infty}$, where $\ket{\psi_{p_{\mathcal{Q}}}}$ is a pure state that is defined on sites $1,2,\ldots,p_{\mathcal{Q}}$. The symmetrization of $\ket{\Psi_{gs}}$ gives the TI state (the uMPS $\ket{\Psi(A)}$). Additionally, the first $p_{\mathcal{Q}}-2$ sites in $\ket{\psi_{p_{\mathcal{Q}}}}$ are not entangled with the last two sites, which ensures that the ground state of the PBC Hamiltonian $H_{\PBC}$ with $p_{\mathcal{Q}}$ sites corresponds to an infinite TI state. In other words, $\ket{\psi_{p_{\mathcal{Q}}}}$ is the ground state of $H_{\PBC}$.

Furthermore, we observe that the Hamiltonian $H$ attaining the quantum limit does not possess ground states with periods other than $p_{\mathcal{Q}}$ and its multiples. Indeed, if $H$ had a ground state with period $p'\neq p_{\mathcal{Q}}$, then the ground-state energy density of $H_{\PBC}$ with $p'$ sites could not exceed the quantum limit. However, our numerical calculations show that, unless $p'$ is a multiple of $p_{\mathcal{Q}}$, the ground-state energy density of $H_{\PBC}$ on $p'$ sites is strictly larger than the quantum limit. This provides strong numerical evidence that $H$ does not possess ground states with periods other than $p_{\mathcal{Q}}$ and its multiples.

\section{Experimental and benchmarking implications}
\label{sec:benchmark}

The 232- and 322-type TI contextuality witnesses studied in this work can be reduced to a $p_{\mathcal{Q}}$-site PBC inequality with the same classical bound and quantum limit. This finite reduction converts what is, a priori, an infinite-system certification problem into a compact, hardware-testable benchmark on a small ring. Operationally, the PBC witness is a translation-averaged sum of few-body correlators and can be interpreted as the energy density of a local ring Hamiltonian, so the test naturally matches the dominant metrology primitive in quantum simulation: estimating local energies rather than reconstructing full states.

This perspective clarifies what makes the finite inequalities compelling on real devices. One does not need tomography, nonlocal measurements, or large-scale entanglement verification; it suffices to estimate a small collection of one- and two-body (and, in some cases, short-range) correlators entering the witness, exactly as in energy-based many-body nonclassicality tests~\cite{j.turaEnergyDetectorNonlocality2017}. In our convention the noncontextual (classical) region is defined by
\begin{equation}
\langle \mathcal{E}_{\PBC} \rangle \ge \mathcal{L},
\end{equation}
so a statistically significant observation of $\langle \mathcal{E}_{\PBC} \rangle < \mathcal{L}$ certifies contextuality.

There are two complementary routes to implementing these finite contextuality benchmarks. The first route uses qubit processors and qubit simulators as a universal front-end by encoding each $d$-level site into a small register of qubits \cite{yangCostLocallyApproximating2025}. In this approach, the qudit observables appearing in the witness (and hence in the effective ring Hamiltonian) are mapped to qubit operators and decomposed into Pauli strings; the witness is then obtained from standard Pauli-basis measurements together with classical post-processing. This makes it straightforward to reuse mature qubit-native toolchains for compilation, error mitigation, and variational energy minimization, while still probing the effective higher-dimensional inequality.

The second route is to implement the local dimension $d>2$ natively, by directly leveraging additional internal energy levels of a single physical particle. Trapped-ion platforms are especially well aligned with this goal, as they support coherent control and readout of multi-level manifolds and have already demonstrated universal processing and entanglement generation beyond the qubit setting~\cite{Ringbauer2022universalQuditProcessor,Hrmo2023nativeQuditEntanglement}. Neutral-atom platforms offer complementary opportunities by exploiting hyperfine, Zeeman, or nuclear-spin degrees of freedom in tweezer arrays, motivating direct qudit-native Hamiltonian engineering and measurement without incurring qubit-encoding overhead~\cite{Henriet2020neutralAtoms,Wang2020quditsHighDim}. From a benchmarking standpoint, native-qudit implementations are particularly attractive here because the witnesses explicitly involve qudit observables acting on higher-dimensional local Hilbert spaces; the test therefore probes whether a device can coherently exploit higher local Hilbert-space dimension as a computational/simulation resource rather than treating extra levels as leakage.

A key practical feature of the present qutrit instances is that both the witness terms and their readout can be expressed using computational-basis population measurements plus local rotations acting only within the $\{|0\rangle,|1\rangle\}$ subspace. In particular, with $\sigma_0=\mathrm{diag}(1,-1,1)$, the second observable can be written as a rotated \emph{diagonal} operator with a different eigenvalue assignment on $|2\rangle$:
\begin{equation}
\begin{aligned}
    &\tilde{\sigma}_0 := |0\rangle\!\langle 0| - |1\rangle\!\langle 1| - |2\rangle\!\langle 2| = \mathrm{diag}(1,-1,-1),\\
& \sigma_1(\theta) = U_{01}(\theta)\,\tilde{\sigma}_0\,U_{01}^\dagger(\theta),
\end{aligned}
\end{equation}
where $U_{01}(\theta)$ acts as a real rotation in the $\{|0\rangle,|1\rangle\}$ subspace and leaves $|2\rangle$ unchanged:
\begin{equation}
U_{01}(\theta)=
\begin{pmatrix}
\cos\theta & \sin\theta & 0\\
-\sin\theta & \cos\theta & 0\\
0 & 0 & 1
\end{pmatrix}.
\end{equation}
Consequently, estimating any correlator built from $\sigma_0$ and $\sigma_1(\theta)$ reduces to applying $U_{01}^\dagger(\theta)$ on the sites where $\sigma_1(\theta)$ is to be measured, followed by standard projective readout in the computational basis; the distinction between $\sigma_0$ and $\tilde{\sigma}_0$ is then implemented entirely in classical post-processing by assigning the appropriate $\pm 1$ eigenvalues to the observed basis state. This is experimentally advantageous because it avoids the need for genuinely non-diagonal joint measurements: all measurement complexity is pushed into calibrated single-site rotations and repeated population readout.

In an experiment, the practical objective is an energy-estimation problem with a certified classical threshold. One chooses a parameterization $H_{\PBC}(\theta)$ of the finite ring Hamiltonian induced by the witness (with $\theta=\theta_0$ saturating the classical bound and $\theta=\theta_1$ attaining the quantum limit), prepares a low-energy state of $H_{\PBC}(\theta)$ using either variational energy minimization~\cite{peruzzoVariationalEigenvalueSolver2014,cerezoVariationalQuantumAlgorithms2021} or an analog/adiabatic preparation when the required couplings are naturally available, and then estimates $\langle \mathcal{E}_{\PBC} \rangle$ by measuring the local terms entering the Hamiltonian density. Scanning $\theta$ provides a controlled interpolation from the noncontextual regime into the contextual regime, while simultaneously allowing one to track how the real-space energy fingerprint sharpens into a $p_{\mathcal{Q}}$-periodic pattern. When true PBC connectivity is unavailable, the same measurements can be performed on a finite open chain by focusing on bulk bond energies; the emergence of the period in the bulk and the convergence of the bulk-averaged energy to the PBC value provide an experimentally practical surrogate for the ring benchmark.

As a concrete illustration, consider the finite ring Hamiltonian obtained from the 322-case study at $p_{\mathcal{Q}}=4$, which can be implemented as a four-site qutrit ring with nearest- and next-to-nearest-neighbor interactions,
\begin{equation}\label{becn:322}
    \begin{aligned}
    H_{\PBC}=\sum_{i=1}^{4} &-3\sigma_0^{(i)} + \sigma_1^{(i)} + \sigma_0^{(i)}\sigma_0^{(i+1)}+\sigma_0^{(i)}\sigma_1^{(i+1)}\\&+\sigma_1^{(i)}\sigma_0^{(i+1)}-\sigma_1^{(i)}\sigma_1^{(i+1)}
    +\sigma_0^{(i)}\sigma_0^{(i+2)}\\&-\sigma_1^{(i)}\sigma_0^{(i+2)}+\sigma_1^{(i)}\sigma_1^{(i+2)},
\end{aligned}
\end{equation}
with
\begin{equation}\footnotesize
	\sigma_0 = \begin{pmatrix}
		1 &  0 & 0\\
		0 & -1 & 0\\
        0 &  0 & 1
	\end{pmatrix},
	\sigma_1(\theta) =
	\begin{pmatrix}
		\cos(2\theta) & -\sin(2\theta)  &0 \\
	   -\sin(2\theta) & -\cos(2\theta)  &0 \\
               0 &         0  &-1
	\end{pmatrix}.
\end{equation}
In this parameterization, $\theta=0$ reproduces the classical-saturating point with ground-state energy density $-3$, while $\theta=\pi/4$ reaches the quantum-optimal point with energy density $-(5+\sqrt{2})/2\approx -3.20711$. Experimentally, these values translate into a direct contextuality certificate: preparing a state whose measured energy density drops below the certified classical threshold $\mathcal{L}$ (with appropriate statistical uncertainty) demonstrates violation of the finite inequality, and measuring the translated local energy contributions provides a simultaneous, device-level diagnostic of the associated $p_{\mathcal{Q}}$-period structure. On platforms where direct analog implementation of all terms in $H_{\PBC}$ is inconvenient, the same certificate can be obtained purely variationally by treating $H_{\PBC}$ as the measured cost function and leveraging the fact that each term is accessible through local $U_{01}(\theta)$ rotations and computational-basis readout.

Beyond certifying contextuality, these finite inequalities serve as targeted benchmarks for higher-dimensional quantum simulation and computation. Recent work has emphasized that qudit-native control can reduce compilation overhead and enable qudit-native simulations of models (including gauge-theory constructions) that are unnatural or inefficient in strictly qubit form, motivating systematic tests of coherent $d>2$ dynamics rather than restricting benchmarks to qubit subspaces~\cite{Wang2020quditsHighDim,Calajo2024su2IonQudits,Zache2023fermionQudit,Gao2023roleofentanglement}. The present PBC inequalities provide such a benchmark in a minimal setting: small system size, few-body measurements, an analytically certified noncontextual threshold, and a direct connection between contextuality violation and an experimentally resolvable real-space period.

\section{Conclusion and outlook}
\label{sec:discussion}

We have studied how translation-invariant contextuality
witnesses, when viewed as local Hamiltonian densities,
organize ground-state structure in infinite one-dimensional
systems. For the 232- and 322-type witness families
considered here, the classical and quantum optima exhibit
sharply different commensurate structures. At the
classical-bound points, the corresponding Hamiltonians
have highly degenerate ground-state sectors supporting
several competing periods. At the quantum-optimal
points, the optimal ground-state sectors are strictly
periodic with enlarged unit cells \(p_{\mathcal{Q}}>1\), so that the
one-site translation symmetry of the Hamiltonian is
spontaneously broken by the extremal ground states. This
mechanism was established analytically in representative
models and observed more broadly across the listed
examples using optimized MPS representatives
and the transfer-matrix period diagnostic.

Beyond its foundational significance, this link elevates our TI contextuality witnesses to practical diagnostics of translation symmetry breaking for correlated quantum matter.  
Because the witnesses involve only few-body correlators and admit a Hamiltonian interpretation, they naturally complement conventional condensed-matter probes of ordering and symmetry breaking.
At the same time, the witness-driven selection of a strict period provides rare analytic control of a symmetry-breaking transition in an infinite 1D setting, including explicit access to the ground-state structure at both the classical and quantum endpoints.

The periodicity enforced by the quantum optimum has direct operational consequences that make the present framework attractive for near-term quantum platforms.
First, the studied TI contextuality witnesses enable finite-size reductions: the infinite-chain TI contextuality problem induces a $p_{\mathcal{Q}}$-site periodic-boundary-condition inequality with the same classical and quantum bounds, and in several cases the resulting finite inequalities are tight.
This yields compact contextuality benchmarks on small rings (e.g., $p_{\mathcal{Q}}=3,4,5$ in our case studies), requiring only local measurements of the few terms appearing in the witness Hamiltonian density \cite{j.turaEnergyDetectorNonlocality2017,Yang2022}.
Such tests can be deployed on digital quantum processors as end-to-end benchmarks of coherent control and many-body state preparation: one prepares an approximate ground state of the $p_{\mathcal{Q}}$-site ring Hamiltonian (e.g., with variational or adiabatic protocols), measures the witness terms, and compares the observed value to the classical bound.
Second, for analog quantum simulators implementing longer open chains, the same witnesses provide experimentally accessible ``energy fingerprints'' in real space: local bond energies exhibit the emergent period and converge under finite-size scaling to the thermodynamic-limit value, enabling cross-checks of both the contextuality certificate and the associated symmetry-broken order.
Finally, since contextuality is a known resource for quantum computational advantage \cite{howardContextualitySuppliesMagic2014,budroniKochenSpeckerContextuality2022}, the present results suggest a concrete many-body design principle for preparing highly contextual resource states: engineer or variationally target phases in which translation symmetry is broken into the period that realizes the strongest witness violation.

Several directions merit further investigation.
One concerns \emph{period selection}: how do the attainable periods and minimal local dimensions depend on witness structure, interaction range, or additional symmetries, and can one systematically design witnesses that target prescribed commensurate orders?
Another is to extend the program beyond 1D. Higher-dimensional analogs could exhibit richer symmetry-breaking patterns, but the classical-versus-quantum extension problem becomes substantially more intricate: LTI is necessary but not sufficient for TI in 2D, and for large local dimension the set of LHV models is no longer semi-algebraic~\cite{wangTwodimensionalTranslationinvariantProbability2018}. Nevertheless, a 2D analog of domino loops—the corner-tiling problem—offers a powerful conceptual tool and has been used extensively in complexity-theoretic studies of Hamiltonian problems~\cite{lagaeAperiodicSetsSquare2006,gottesmanQuantumClassicalComplexity2009,cubittUndecidabilitySpectralGap2015,wangTwodimensionalTranslationinvariantProbability2018}.
Addressing
these questions may clarify how far the connection
between contextuality, energy-density witnesses, and
spatial ordering extends beyond the one-dimensional
models analyzed here.

\section*{Acknowledgments}
This work was supported by the NSFC (No.~12574536). The authors acknowledge Miguel Navascu\'{e}s and Paolo Abiuso for valuable discussions.

\section*{Author contribution}
Z.W. and X.Z. contributed to the conceptual formulation of this work. 
X.Z., K.Y. and L.Z. contributed to the
design and analysis of the numerical simulations. 
All authors contributed to discussion of the results and text writing. 
Large language models assisted with text refinement, critical feedback, and reference checking. All scientific claims, derivations, numerical results, and the final text were independently verified and approved by the authors.

\bibliographystyle{quantum}
\bibliography{ref}

\appendix

\section{The analytical solution of the 232-type contextuality witness} \label{appendix:232analytical}
\subsection{The optimal observables and state}
The 232-type contextuality witness is in the form of \begin{equation}\label{appendix-232-ineq}
   \langle \mathcal{E} \rangle_{\text{TI}} = \langle \sigma^{(1)}_1\sigma^{(2)}_0+\sigma^{(1)}_1\sigma^{(2)}_1-\sigma^{(1)}_2\sigma^{(2)}_0+\sigma^{(1)}_2\sigma^{(2)}_1 \rangle_{\text{TI}}  \geq -2.
\end{equation}
Using MPS-based optimization and NPA, we found that the maximal quantum violation can be saturated when $d=5, D=4$. 

The simplest optimal observables we found for Eq.~\eqref{appendix-232-ineq} are 
\begin{equation}\label{232ob}{\footnotesize
\begin{aligned}
    &\sigma_0 =     \begin{pmatrix*}[r]
        1 & 0 & 0 & 0 & 0\\
        0 & 1 & 0 & 0 & 0\\
        0 & 0 & -1 & 0 & 0\\
        0 & 0 & 0 & 1 & 0\\
        0 & 0 & 0 & 0 & 1
    \end{pmatrix*},   \sigma_1 = -     \begin{pmatrix*}[r]
        0 & 0 & 1 & 0 & 0\\
        0 & 1 & 0 & 0 & 0\\
        1 & 0 & 0 & 0 & 0\\
        0 & 0 & 0 & 0 & 1\\
        0 & 0 & 0 & 1 & 0
    \end{pmatrix*},
       \sigma_2 =     \begin{pmatrix*}[r]
        1 & 0 & 0 & 0 & 0\\
        0 & 1 & 0 & 0 & 0\\
        0 & 0 & 1 & 0 & 0\\
        0 & 0 & 0 & 1 & 0\\
        0 & 0 & 0 & 0 & -1
    \end{pmatrix*}.
\end{aligned}}
\end{equation}

The uniform MPS tensor $\{A^s\}$ (in right orthogonal form) is
\begin{equation}\footnotesize{
    \begin{aligned}
    & A^0 = \begin{pmatrix*}[r]
         0&        0&       0&  0\\
         0&        0&       0&  0\\
         a_1&  -a_1&  0&  0\\
         a_2&   -a_2&   0&  0
    \end{pmatrix*} ,
     A^1 = \begin{pmatrix*}[r]
          \frac{1}{2}&   \frac{1}{2}&  0&  0\\
         -\frac{1}{2}&  -\frac{1}{2}&  0&  0\\
          0&   0&  0&  0\\
          0&   0&  0&  0
    \end{pmatrix*},\\& A^2 = \begin{pmatrix*}[r]
          0&        0 &      0&  0\\
          0   &     0  &     0&  0\\
         -a_2&    a_2&   0&  0\\
          a_1&  -a_1&  0&  0
    \end{pmatrix*},
     A^3 = \begin{pmatrix*}[r]
         0&  0&  -b_2&  b_1\\
         0&  0&  -b_2&  b_1\\
         0&  0&  0&  0\\
         0&  0&  0&  0
    \end{pmatrix*},\\& A^4 = \begin{pmatrix*}[r]
         0&  0&  b_1&  b_2\\
         0&  0&  b_1&  b_2\\
         0&  0&  0&  0\\
         0&  0&  0&  0
    \end{pmatrix*}, \\ 
\end{aligned}}
\end{equation}

where $a_1=-0.38201775, a_2=0.59503147,a_1^2+a_2^2=1/2, b_1=0.28534978, b_2=0.41057947,b_1^2+b_2^2=1/4$. Note that we still have a gauge freedom in the form of \begin{equation}
    U=\begin{pmatrix*}[r]
         \mathbb{I}_{2}  &  \mathbf{0}\\
         \mathbf{0} &   u
    \end{pmatrix*}.
\end{equation}
We use $u$ to transform $(b_1,b_2)$ into $(1/2,0)$, and $(a_1,a_2)$ becomes $(0.27001975,0.65352073)$. Therefore, the analytical solution of $\{A^s\}$ depends only on the parameter $a_1$. 

Interestingly, the transfer matrix $E_A(X)=\sum_{i}A^iX(A^i)^{\dagger}$ is independent of  parameter $a_1$, and it has a simple form of
\begin{equation}\footnotesize
    E_A=\begin{pmatrix*}[r]
  \frac{1}{4}&  \frac{1}{4}& 0& 0&  \frac{1}{4}&  \frac{1}{4}& 0& 0& 0& 0& \frac{1}{2}& 0& 0& 0& 0& \frac{1}{2}\\
 -\frac{1}{4}& -\frac{1}{4}& 0& 0& -\frac{1}{4}& -\frac{1}{4}& 0& 0& 0& 0& \frac{1}{2}& 0& 0& 0& 0& \frac{1}{2}\\
  0&  0& 0& 0&  0&  0& 0& 0& 0& 0& 0& 0& 0& 0& 0& 0\\
  0&  0& 0& 0&  0&  0& 0& 0& 0& 0& 0& 0& 0& 0& 0& 0\\
 -\frac{1}{4}& -\frac{1}{4}& 0& 0& -\frac{1}{4}& -\frac{1}{4}& 0& 0& 0& 0& \frac{1}{2}& 0& 0& 0& 0& \frac{1}{2}\\
  \frac{1}{4}&  \frac{1}{4}& 0& 0&  \frac{1}{4}&  \frac{1}{4}& 0& 0& 0& 0& \frac{1}{2}& 0& 0& 0& 0& \frac{1}{2}\\
  0&  0& 0& 0&  0&  0& 0& 0& 0& 0& 0& 0& 0& 0& 0& 0\\
  0&  0& 0& 0&  0&  0& 0& 0& 0& 0& 0& 0& 0& 0& 0& 0\\
  0&  0& 0& 0&  0&  0& 0& 0& 0& 0& 0& 0& 0& 0& 0& 0\\
  0&  0& 0& 0&  0&  0& 0& 0& 0& 0& 0& 0& 0& 0& 0& 0\\
  \frac{1}{4}& -\frac{1}{4}& 0& 0& -\frac{1}{4}&  \frac{1}{4}& 0& 0& 0& 0& 0& 0& 0& 0& 0& 0\\
  0&  0& 0& 0&  0&  0& 0& 0& 0& 0& 0& 0& 0& 0& 0& 0\\
  0&  0& 0& 0&  0&  0& 0& 0& 0& 0& 0& 0& 0& 0& 0& 0\\
  0&  0& 0& 0&  0&  0& 0& 0& 0& 0& 0& 0& 0& 0& 0& 0\\
  0&  0& 0& 0&  0&  0& 0& 0& 0& 0& 0& 0& 0& 0& 0& 0\\
  \frac{1}{4}& -\frac{1}{4}& 0& 0& -\frac{1}{4}&  \frac{1}{4}& 0& 0& 0& 0& 0& 0& 0& 0& 0& 0 
    \end{pmatrix*}.
\end{equation}
The nonzero eigenvalues of $E_A$ are $\{e^{\frac{2\pi ik}{3}}:k=0,1,2\}$, which means that the uMPS $\ket{\psi(A)}$ is not injective. 
Since $E_A$ is independent of $a_1$, so are the left and right fixed points, which are \begin{equation}\footnotesize
    l=\begin{pmatrix*}[r]
         \frac{1}{3}&  0&  0&  0\\
         0&  \frac{1}{3}&  0&  0\\
         0&  0&  \frac{1}{6}&  0\\
         0&  0&  0&  \frac{1}{6}
    \end{pmatrix*}, r=\begin{pmatrix*}[r]
         1&  0&  0&  0\\
         0&  1&  0&  0\\
         0&  0&  1&  0\\
         0&  0&  0&  1
    \end{pmatrix*}.
\end{equation}
The energy density $e$ can thus be written as 
\begin{equation}
\begin{tikzpicture}[
    every node/.style={font=\small},
    tensor/.style={
        draw, rounded corners,
        minimum width=0.42cm,
        minimum height=0.42cm,
        inner sep=1pt
    },
    env/.style={draw, circle, inner sep=1pt},
    ham/.style={
        draw, rectangle,
        minimum width=1.35cm,
        minimum height=0.34cm,
        inner sep=1pt
    }
]


\node at (0.35,0) {$e=$};
\node[env] (l) at (0.8,0) {$l$};

\node[tensor] (m1) at (1.55, 0.55) {$A$};
\node[tensor] (m2) at (2.55, 0.55) {$A$};
\node[tensor] (n1) at (1.55,-0.55) {$\bar A$};
\node[tensor] (n2) at (2.55,-0.55) {$\bar A$};

\node[env] (r) at (3.3,0) {$r$};
\node[ham] (H) at (2.05,0) {$h$};

\draw (m1) -- (m2);
\draw (n1) -- (n2);

\draw (l) to[out=70,in=180] (m1);
\draw (l) to[out=-70,in=180] (n1);
\draw (r) to[out=110,in=0] (m2);
\draw (r) to[out=-110,in=0] (n2);

\draw (m1.south) -- (m1.south |- H.north);
\draw (m2.south) -- (m2.south |- H.north);
\draw (n1.north) -- (n1.north |- H.south);
\draw (n2.north) -- (n2.north |- H.south);

\end{tikzpicture}
\end{equation}
Now, $e$ is a function that only depends on $a_1$: \begin{equation}
    e=-\frac{2}{3} \left(-4 a_1^2+2a_1 \sqrt{2-4 a_1^2} +3\right).
\end{equation}
The optimal value of $a_1$ is $\sqrt{\frac{1}{8} \left(2-\sqrt{2}\right)}$, and the optimal value of $e$ is $-\frac{2}{3} \left(2+\sqrt{2}\right)\approx -2.276142374915397$, which matches the MPS value and NPA value to ten significant digits.




\subsection{Incorporating LTI and symmetries in the NPA hierarchy}\label{appendix-npa232}
The goal in this subsection is to derive the tight lower bound of Eq.~\eqref{appendix-232-ineq}. To compute the tight lower bound, we use NPA, which is a convergent hierarchy of SDPs characterizing the set of quantum correlations. The tricky part of Eq.~\eqref{appendix-232-ineq} is that the operators $\{\sigma_x^{(1)}\sigma_y^{(2)}\}$ are evaluated on the two-body reduced states of infinite TI states. 
By TI we mean \begin{equation}
    \rho_{(1,2,\ldots,n)} = \rho_{(1+k,2+k,\ldots,n+k)}\quad \text{for  } k\in \mathbb{Z}, n\in \mathbb{N}, 
\end{equation}
where $\rho_{(1,2,\ldots,n)}$ is the reduced state of the infinite TI state $\rho$ on sites $(1,2,\ldots,n)$.
However, it is not possible to characterize the set of TI reduced states \cite{blakajSetReducedStates2024}.
Nevertheless, we can relax the set of reduced TI states to the set of LTI states to derive lower bounds on the maximal violation. 
By saying $\rho_n$ is $n$-LTI we mean \begin{equation}
    \rho_{(1,2,\ldots,n-1)} = \rho_{(2,3,\ldots,n)}.
\end{equation}
The relaxation of $n$-LTI is tighter when $n$ is larger.
As $n\to \infty$, the relaxation becomes exact. The construction of LTI-NPA is given in Section~\ref{subsec:lti-npa}.
However, as $n$ grows, the size of the SDP grows exponentially. 
Therefore, it is crucial to exploit symmetries and LTI to simplify the problem \cite{ioannouNoncommutativePolynomialOptimization2022}.

\subsubsection{Symmetries}
We first briefly introduce the symmetries in the NPA hierarchy (a detailed methodological introduction can be seen in \cite{ioannouNoncommutativePolynomialOptimization2022}). Symmetries in contextuality witnesses are obtained by relabeling the inputs (outputs) of each player and permuting players \cite{DenisThesis}.
Relabeling the inputs of each player corresponds to the permutation group $S_3$ on $\{0,1,2\}$. A permutation $\pi\in S_3$ acts on Eq.~\eqref{appendix-232-ineq} in the form of 
\begin{equation}\label{eq:projectors}
    \sigma_x^{(1)} \xrightarrow{\pi} \sigma_{\pi(x)}^{(1)},
    \sigma_y^{(2)} \xrightarrow{\pi} \sigma_{\pi(y)}^{(2)}.
\end{equation}
Relabeling the outputs of each player corresponds to sign symmetries (it can be readily checked by Eq.~\eqref{eq:projectors}), i.e., 
\begin{equation}
    \sigma_x^{(1)} \xrightarrow{} -\sigma_x^{(1)},
    \sigma_y^{(2)} \xrightarrow{} -\sigma_y^{(2)}.
\end{equation}
Permuting players corresponds to 
\begin{equation}
    \sigma_x^{(1)} \xrightarrow{} \sigma_x^{(2)} ,
    \sigma_y^{(2)} \xrightarrow{} \sigma_y^{(1)} .
\end{equation}
If there are more parties, the symmetry group $G_{\mathcal{L}}$ of the inequalities will consist of more elements. 
The symmetry group of the optimization problem, denoted by $G_{\mathcal{E}}$, is the set of symmetries that leaves the objective $\mathcal{E}$ invariant, i.e., \begin{equation}
    G_{\mathcal{E}}=\{g\in G_{\mathcal{L}}|g(\mathcal{E})=\mathcal{E}\}
\end{equation} 
Once $G_{\mathcal{E}}$ is determined, we can symmetrize the basis $(M_k)_k$ using $G_{\mathcal{E}}$:
\begin{equation}
    \overline{M}_k= \frac{1}{|G_{\mathcal{E}}|}\sum_{g\in G_\mathcal{E}}g(M_k).
\end{equation}
Consequently, we have $\overline{x}_k=\text{tr}(\rho \overline{M}_k)$. 
After symmetrization, there may be redundancies in $(\overline{M}_k)_k$. We choose a basis $(\widetilde{M}_i)_i$ ($i=1,2,\ldots,s'$) from $(\overline{M}_k)_k$ and set $\widetilde{x}_i=\text{tr}(\rho \widetilde{M}_i)$ as the new variables. Since $g(\mathcal{E})=\mathcal{E}$, the following SDP has the same optimal value as Eq.~\eqref{npa-org}: \begin{equation}
	\label{sdp:npasymmetrized}
	\begin{aligned}
		\text{min} \quad &\sum_{i=1}^{s'} \widetilde{c}_{i} \widetilde{x}_i \\
		\text{s.t.}\quad &   \widetilde{x}_1=1,  \textstyle\sum_{i=1}^{s'} \widetilde{A}_{i} \widetilde{x}_i \succeq 0,\\
	\end{aligned}
\end{equation}
where $\widetilde{c}_i, \widetilde{A}_i$ are adjusted with respect to $(\widetilde{M}_i)_i$.
By eliminating redundancies, $s'$ can be substantially smaller than $s$.

The SDP Eq.~\eqref{sdp:npasymmetrized} can be further simplified. We compute the linear representation $\rho$ of $G_{\mathcal{E}}$ in the generating sequence $(\mathcal{Q}_i)_i$, that is, \begin{equation}
    g(Q)=\rho_g(Q).
\end{equation}
The representation $\rho$ can be decomposed into the direct sum of irreducible representations, enabling a block diagonalization of $\widetilde{A}_i$. In other words, using group representation theory, we can find a matrix $P$ such that for all $i\in\{1,2,\ldots,s'\}$, $P\widetilde{A}_iP^{-1}$ share the same block diagonal structure. 
Let $k_1,k_2,\ldots,k_r$ be the dimension of the diagonal blocks; then $\widetilde{A}_i$ can be represented by \begin{equation}\small
P\widetilde{A}_iP^{-1}=
    \begin{pmatrix}
        \widetilde{A}_{i,k_1} & \\
                  &  \ddots\\
                  &&\widetilde{A}_{i,k_r}
    \end{pmatrix}.
\end{equation} The original constraint $\textstyle\sum_{i=1}^{s'} \widetilde{A}_{i} \widetilde{x}_i \geq 0$ becomes \begin{equation}
   \textstyle\sum_{i=1}^{s'} \widetilde{A}_{i,k_j}\widetilde{x}_i\geq 0 \text{ for all } j\in\{1,\ldots,r\}.
\end{equation}
The final SDP is given by \begin{equation}
	\label{sdp:npasymmetrizedbd}
	\begin{aligned}
		\text{min} \quad &\sum_{i=1}^{s'} \widetilde{c}_{i} \widetilde{x}_i \\
		\text{s.t.}\quad &   \widetilde{x}_1=1, 
         \textstyle\sum_{i=1}^{s'} \widetilde{A}_{i,k_j}\widetilde{x}_i\geq 0 \text{ for all } j\in\{1,\ldots,r\}.\\
	\end{aligned}
\end{equation}

Using Eq.~\eqref{eq:lti}, it is straightforward to add LTI constraints in Eq.~\eqref{sdp:npasymmetrizedbd}, and we denote those constraints by $L'\tilde{x}=0$. The LTI-NPA using symmetries is then given by \begin{equation}
	\label{sdp:lti-npasymmetrizedbd}
	\begin{aligned}
		\text{min} \quad &\sum_{i=1}^{s'} \widetilde{c}_{i} \widetilde{x}_i \\
		\text{s.t.}\quad &   \widetilde{x}_1=1, L'\tilde{x}=0,
         \textstyle\sum_{i=1}^{s'} \widetilde{A}_{i,k_j}\widetilde{x}_i\geq 0 \text{ for all } j\in\{1,\ldots,r\}.\\
	\end{aligned}
\end{equation}
Although symmetries can significantly simplify the optimization problem, the side effect is that the power of LTI constraints may be weakened. The reason is that after symmetrization, if the number of terms in $(\widetilde{M}_i)_i$ that satisfy Eq.~\eqref{eq:lti} is reduced, the optimal value of Eq.~\eqref{sdp:lti-npasymmetrizedbd} may be worse than that of Eq.~\eqref{lti-npa}. To exploit the trade-off between symmetries and LTI, we need to identify which kinds of symmetries are compatible with LTI. 

\subsubsection{SDP simplifications from symmetry for the 232-type inequality}
The symmetry group $G_{\mathcal{E}}$ of Eq.~\eqref{appendix-232-ineq} consists of $32$ elements. The generators of $G_{\mathcal{E}}$ are 
\begin{enumerate}
    \item  $\sigma^{(1)}_0 \rightarrow - \sigma^{(1)}_0$,
        \item $\sigma^{(n)}_2 \rightarrow - \sigma^{(n)}_2$,
    \item $\sigma^{(1)}_1 \rightarrow  \sigma^{(1)}_2,\sigma^{(1)}_2 \rightarrow  \sigma^{(1)}_1,\sigma^{(2)}_0 \rightarrow - \sigma^{(2)}_0$,
    \item $\sigma^{(n)}_0 \rightarrow  \sigma^{(n)}_1,\sigma^{(n)}_1 \rightarrow  \sigma^{(n)}_0,\sigma^{(n-1)}_2 \rightarrow  -\sigma^{(n-1)}_2$,
    \item $\sigma^{(1)}_1 \rightarrow  -\sigma^{(1)}_2,\sigma^{(1)}_2 \rightarrow -\sigma^{(1)}_1,\sigma_1^{(2)}\rightarrow -\sigma_1^{(2)},\sigma_2^{(2)}\rightarrow -\sigma_2^{(2)}, \sigma_{x}^{(i)}\rightarrow - \sigma_{x}^{(i)} \text{ for } x\in\{0,1,2\},i\in\{3,\ldots,n\}$.
\end{enumerate}

When $n=3$, we use the generating sequence $\mathcal{Q}_3$. $\mathcal{Q}_3$ consists of $64$ elements and the basis $(M_k)_k$ contains $1297$ elements. There are $117$ pairs of LTI terms in $x$. After symmetrization, there are only $51$ elements in $(\widetilde{M}_k)_k$. The number of variables is significantly reduced. However, there are only $2$ pairs of LTI terms left in $\widetilde{M}_k$. The optimal value we get from $(\widetilde{M}_i)_i$ is almost the same as that obtained without imposing the LTI constraints, which is much worse than that of $({M}_k)_k$. This means that some symmetries are not compatible with LTI.

In order to exploit advantages of LTI, we use a subgroup of $G_{\mathcal{E}}$ instead. Sign symmetries are friendly to LTI, because during symmetrization, each $M_k$ goes to zero or stays unchanged. Let $G_{s} \subset G_{\mathcal{E}}$ be the sign symmetry group and $g\in G_{s}$, then we have $g(M_i)=M_i$ or $g(M_i)=-M_i$. If $g(M_i)=M_i$ for all $g\in G_s$, we have $\frac{1}{|G_s|}\sum_{g\in G_s} g(M_k)=M_k$. If $g(M_i)=-M_i$ for some $g\in G_s$, we have $\frac{1}{|G_s|}\sum_{g\in G_s} g(M_k)=0$. Symmetrization using $G_{s}$ may still lose some LTI terms, but we find that the optimal value is close to the original optimization problem. The sign symmetry group $G_s$ consists of $4$ elements. After symmetrization using $G_s$, there are $326$ elements in $(\widetilde{M}_k)_k$ and $28$ pairs of LTI terms. Although the number of variables is larger, the resulting bound is tighter. Other types of symmetries are less compatible with LTI, as the symmetrization process transforms each $M_k$ into a linear combination of $\sigma_x^{(i)}$, which significantly deviates from the original form of $M_k$.

For $n=5$, we choose the generating sequence $\mathcal{Q}_5$. The lower bound obtained from LTI5-NPA5 is $-2.276142374$. This value agrees with the MPS simulation, thus validating the $232$-type model that achieves the maximal quantum violation.

\end{document}